\documentclass[journal,11pt,doublespace,onecolumn]{IEEEtran}
\usepackage{bm}
\usepackage{amsfonts}
\usepackage{mathtools}
\usepackage{graphicx}
\usepackage{epstopdf}
\DeclareGraphicsExtensions{.eps}
\usepackage{url}
\usepackage{color}
\usepackage{balance}
\usepackage{float}

\def\T{{ \mathrm{\scriptscriptstyle T} }}
\def\P{{ \mathrm{pr} }}
\def\v{{\varepsilon}}

\def\de{\overset{\Delta}{=}}

\DeclarePairedDelimiterX{\norm}[1]{\lVert}{\rVert}{#1}

\newtheorem{remark}{Remark}
\newtheorem{theorem}{Theorem}%\newtheorem{theorem}{Theorem}[section]

\newtheorem{assumption}{Assumption}

\usepackage{cite}
\ifCLASSINFOpdf
\else
\fi
\usepackage{amsmath}
\begin{document}
\title{SLANTS: Sequential Adaptive Nonlinear Modeling of Vector Time Series}

\author{Qiuyi~Han, %~\IEEEmembership{Student Member,~IEEE,}
		Jie~Ding, %~\IEEEmembership{Student Member,~IEEE,}
        Edoardo~Airoldi, %~\IEEEmembership{Member,~IEEE,}
        and~Vahid~Tarokh,~\IEEEmembership{Fellow,~IEEE}% <-this % stops a 
\thanks{This work is supported by Defense Advanced Research Projects Agency (DARPA) grant numbers W911NF-14-1-0508 and  N66001-15-C-4028.}
%\thanks{Q.~Han, J.~Ding, E.~Airoldi, and V. Tarokh are with the John A. Paulson School of Engineering and Applied Sciences, Harvard University, Cambridge, MA, 02138.}% <-this % stops a space
}%



\maketitle

\begin{abstract}
We propose a method for adaptive nonlinear sequential modeling of vector-time series  data. 
Data is modeled as a nonlinear function of past values corrupted by noise, and the underlying  non-linear function is assumed to be approximately expandable in a spline basis.
We cast the modeling of data as finding a good fit representation in the  linear span of multi-dimensional spline basis, and use  a variant of  $l_1$-penalty regularization
 in order to reduce the dimensionality of representation. Using adaptive filtering techniques, we design our online algorithm to automatically tune the underlying parameters based on the minimization of the regularized sequential prediction error. 
We demonstrate the generality and flexibility of the proposed approach on both synthetic and real-world datasets. Moreover, we analytically investigate the performance of our algorithm by obtaining both bounds of the prediction errors, and consistency results for variable selection.
\end{abstract}


% \fi
\IEEEpeerreviewmaketitle

\section{Introduction} \label{sec:intro}
\IEEEPARstart{S}{equentially} observed vector-time series are emerging in various applications. In most these applications modeling nonlinear functional inter-dependency between present and past data is crucial for both representation and prediction. %Different from the case of independent and identically distributed (i.i.d.) data, a major task in time series analysis is to develop a  (stochastic) forecasting model that is able to predict the future by exploring the past.  
%In developing a reliable and flexible model that can be widely used in practical scenarios, a practitioner is often faced with these challenges: 1) sequential inference, which means that the inference and prediction at each time step should be made by using only the data before that time, and be updated upon each newly arrived data; in view of that, analysis techniques such as trend-cycle decomposition that lacks such sequential nature may not be applicable; 2) nonlinearity, in which the functional relation is not necessarily linear, and classical linear approaches such as autoregressive-moving-average model \cite{anderson2011statistical,box2011time,brockwell2013time} may not suffice; 3) adaptivity, which means that the data-generating model is varying over time, and a Kalman filter-type update is desired to track new environments; 4) high dimensionality, which naturally arises when a practitioner seeks to look for more candidate models as sample size grows. 
This is a challenging problem given that often in various applications fast online implementation, adaptivity and ability to handle high dimensions are basic requirements for nonlinear modeling.  For example, environmental science combines high dimensional weather signals for real time prediction \cite{xingjian2015convolutional}. In epidemics, huge amount of online search data is used to form fast prediction of influenza epidemics \cite{yang2015accurate}. In finance, algorithmic traders demand adaptive models to accommodate a fast changing stock market. In robot autonomy, there is the challenge of learning the high dimensional movement systems \cite{vijayakumar2005incremental}. 
These tasks usually take high dimensional input signals which may contain a large number of irrelevant signals. % and require fast sequential learning
In all these applications, clearly methods to remove redundant signals, and learn the nonlinear model with low computational complexity are well sought after.
This motivates our work in this paper, where we propose an approach to sequential nonlinear adaptive modeling of potentially high dimensional vector time series.

Inference of nonlinear models has been a notoriously difficult problem, especially for large dimensional data%$L$
\cite{friedman1991multivariate, vijayakumar2005incremental,huang2010variable}.
In low dimensional settings, there have been remarkable parametric and nonparametric nonlinear time series models that have been applied successfully to data from various domains. Examples include threshold models\cite{tong2012threshold}, generalized autoregressive conditional hetero-scedasticity models\cite{gourieroux2012arch}, multivariate adaptive regression splines (MARS)\cite{friedman1991multivariate}, generalized additive models\cite{hastie1990generalized}, functional coefficient regression models\cite{cai2000functional}, etc. 
However, some of these methods may suffer from prohibitive computational complexity.  Model selection using some of these approaches is yet another challenge as they may not eliminate insignificant predictors.    
In contrast, there exist high dimensional nonlinear time series models (\ref{model:00}) that are mostly inspired by high dimensional statistical methods. 
%They typically fall into two categories.
In one approach,   a small subset of significant variables is first selected and then nonlinear time series models are applied to selected variables. For example, independence screening techniques such as \cite{zhang2009identifiability,zhang2011kernel,fan2012nonparametric} or the %remarkable 
MARS may be used to do variable selection. 
In another approach,  dimension reduction method such as least absolute shrinkage and selection operator (LASSO) \cite{tibshirani1996regression} are directly applied to nonlinear modeling. 
Sparse additive models have been developed in recent works of Ravikumar \cite{ravikumar2009sparse} and Huang\cite{huang2010variable}.  These approaches seem to be very promising, and may benefit from additional  reductions in computational complexity.

 %applies group Lasso \cite{yuan2006model} on nonparametric additive models. %Because nonparametric additive models turn the nonlinear problem into a linear regression by approximating $f$ with a linear combination of splines, the standard dimension reduction tools such as Lasso naturally applies. The ``group'' is used to impose a natural constraint that the set of coefficients corresponding to a nonlinear function are either zero or nonzero at the same time.
%
%In practice when the data are sequentially obtained, it is appealing to develop a sequential inference procedure. 
%In this work, we build upon the idea of the second category above as it is more suitable in a changing environment.
In this work, we will build on the  latter category and develop an adaptive and online model for nonlinear modeling.  Our method is sequential which provides computational benefits as we avoid applying batch estimation up on sequential arrival of data.  It also provides resilience to time-variation of data, which may be important as in many practical applications \cite{xingjian2015convolutional,yang2015accurate,vijayakumar2005incremental}, the functional dependency between present and past data seems to be time-varying.  In fact, it is widely believed that a robust inference procedure must be adaptive to new environments (data generating processes).  However, it is common to assume that these time-variations are smooth \cite{babadi2010sparls}, and this assumption will be made in the sequel.  Using this smoothness assumption,  we present a Sequential Learning Algorithm for Nonlinear Time Series (SLANTS). 
Specifically, we will use the spline basis to dynamically approximate the nonlinear functions. As common in adaptive filtering, we give a larger weights to more recent data points. We use  group LASSO for dimensionality reduction in
our simultaneous estimation and model selection, and for sequential update.  To this end, we re-formulate our group LASSO regularization into a recursive estimation problem that produces an estimator close to the maximum likelihood estimator from batch data. 

The outline of this paper is given next.
In Section \ref{sec:SLANTS}, we formulate the problem mathematically and present our inference algorithm. In Section \ref{sec:theory}, we present our theoretical results regarding prediction error and model consistency. In Section \ref{sec:Num_Results}, we provide numerical results using both synthetic data and two sets of real data examples. The results demonstrate excellent performance of our methods.% Finally, we will make our conclusions in Section \ref{sec:conclusion}.

\section{Sequential modeling of nonlinear time series} \label{sec:SLANTS}
In this section, we first present our mathematical model and cast our problem as $l_1$-regularized linear regression . We then propose an EM type algorithm to sequentially estimate the underlying coefficients. Finally we disclose methods for tuning the underlying parameters. Combining our proposed EM estimation method with automatic parameter tuning, we tailor our algorithm to sequential vector time series applications.
\subsection{Formulation of SLANTS}
Consider a multi-dimensional vector time series given by 
\[
\bm X_t=[X_{1,t}, \ldots, X_{D,t}]^\T \in \mathbb{R}^D, t = 1,2,\ldots
\] 
Our main objective in this paper is to predict the value of $\bm X_T$ at time $T$ given the past observations $\bm X_{T-1},\ldots, \bm X_{1}$. Without loss of generality, for simplicity we present our results for the prediction of scalar random variable $X_{1,T+1}$. Let $\bm X_{t, -j} = [X_{1,t}, \ldots, X_{j-1,t},X_{j+1,t}, \ldots, X_{D,t}]^\T.$ %denote the $(D-1) \times 1$ vector that takes away $X_{t,j}$ from $\bm X_t$. 
We start with a general formulation 
\begin{align} \label{model:0}
  X_{1,T} = f(\bm X_{T,-1}, \bm X_{T-1},\ldots,\bm X_{T-L})+\v_T
\end{align}
where $f(\cdot,  \cdots, \cdot)$ is smooth (or at least piece-wise smooth),  $\v_t$ are independent and identically distributed (i.i.d.) zero mean %unit variance 
random variables and the lag order $L$ is a finite but unknown nonnegative integer.

This general formulation encompasses  three special but important cases.  The first case is when $L=0$,  and there is only instantaneous functional relationship among random variables:
%\begin{align} \label{model:space}
$ X_{1,T} = f(\bm X_{T,-1})+\v_T$
%\end{align} 
and this is just a standard regression problem.
The second  case is when $D=1$, and the Equation  (\ref{model:0}) reduces to the one-dimensional (non)linear autoregressive model.
%\begin{align} \label{model:time}
%$ X_{T} = f(X_{T-1},\ldots,X_{T-L})+\v_T.$
%\end{align}   
%
The third case is when $f(\cdot, \cdots, \cdot)$ includes variables related to only times before $t$.
\begin{align} \label{model:pred}
  X_{1,T} = f(\bm X_{T-1},\ldots,\bm X_{T-L})+\v_T .
\end{align} 
This case is of practical interest for prediction purposes.
We rewrite the model in (\ref{model:pred}) as 
\begin{align} 
X_{1,T} &= f(X_{1,T-1},\ldots,X_{1,T-L},\ldots,X_{D,T-1},\ldots,X_{D,T-L}) +\v_T . \nonumber
\end{align}
%where $\tilde{D} = DL$ and $\{\tilde{X}_t\}$ is a one-dimensional time series formed by interweaving $\{\bm X_{1,t}\},\ldots,\{\bm X_{D,t}\}$. 
%With a slight abuse of notations, we rewrite the above model by
%\begin{align}\label{model:00}
%  Y_{T} = f(\bm X_{T-1}, \ldots,\bm X_{T-L}) + \v_T
%\end{align}
%where $\{X_t\},\{Y_t\} $ are two one-dimensional time series that are sequentially observable.
With a slight abuse of notation, we rewrite the above model as 
\begin{align} \label{model:00}
  Y_T
  %&
  = f(X_{1,T},\ldots,X_{\tilde{D},T})+\v_T
\end{align}
with observations $Y_T=X_{1,T}$ and $[X_{1,T},\ldots,X_{\tilde{D},T}]$ $=$ $[X_{1,T-1},\ldots, X_{1,T-L}, \ldots, X_{D,T-1} , \ldots, X_{D,T-L}]$, where $\tilde{D} = DL$. 
%If $f$ is a linear function, (\ref{model:00}) can be treated as a standard linear regression problem.\\
%
To estimate $f(\cdot,  \cdots, \cdot)$, we consider the following least squares formulation
\begin{align} \label{eq1}
  \min_{f} \sum_{t=1}^T w_{T,t} (Y_{t} - f(X_{1,t}, \ldots,X_{\tilde{D},t}))^2 
\end{align}
where $\{w_{T,t} \in [0,1]\}$ are weights used to emphasize varying influences of the past data. The appropriate choice of $\{w_{T,t} \in [0,1]\}$ will be later discussed in section  \ref{subsec:tuningParams}.
  
In order to estimate the nonlinear function $f(\cdot,  \cdots, \cdot)$ , we further assume a nonlinear additive model, i.e. 
\begin{align}
  &f(X_{1,t}, \ldots,X_{\tilde{D},t}) = \mu+\sum_{i=1}^{\tilde{D}} f_i(X_i) , \quad
  E\{ f_i(X_i) \}=0, \label{eq5}
\end{align}
where $f_i$ are  scalar functions, and expectation is with respect to the stationary distribution of $X_i$. The second condition is required for identifiability. 
To estimate $f_i$, we use B-splines (extensions of polynomial regression techniques \cite{wahba1990spline}). 
%The commonly used spline basis are B-splines and truncated power splines. Each $f_i$ can be approximated by splines.  
%For the reason that will be clear later, we use B-spline basis.
%We use B-spline in this paper.
In our presentation, for brevity, we consider the additive model mainly but note that our methods can be extended to models where there exist interactions among $\bm X_1,\ldots, \bm X_{\tilde{D}}$ using multidimensional splines in a straight-forward manner. 

Incorporating the B-spline basis into regression, we write 
\begin{align} \label{eq:spline}
f_i(x) = \sum_{j=1}^v c_{i,j} b_{i,j}(x), \quad 
b_{i,j}(x) = B(x \mid t_{i,1},\ldots,t_{i,v}), 
\end{align}
where $t_{i,1},\ldots,t_{i,v}$ are the knot sequence and $b_{i,j}$ are the coefficients associated with the B-spline basis.  
Here, we have assumed that there are $v$ spline basis of degree $k$ for each $f_i$.
Replacing these into (\ref{eq1}),  the problem of interest is now the minimization of    
\begin{align} \label{eq3}
  \hat{e}_T = \sum_{t=1}^T w_{T,t} \biggl\{Y_{t} - \mu - \sum_{i=1}^{\tilde{D}} \sum_{j=1}^v c_{i,j} b_{i,j}(X_{i,t}) \biggr\}^2 
\end{align}
over $c_{i,j}, \, i=1,\ldots,\tilde{D}, \, j=1,\ldots,v$, under the constraint %(imposed by  $E\{ f_i(X_{i}) \}=0$ in (\ref{eq5}) ):
\begin{align}
\sum_{t=1}^T  \sum_{j=1}^v c_{i,j} b_{i,j}(x_{i}) = 0, \mbox{ for } i=1,\ldots,L
\end{align}
which is the sample analog of the constraint in (\ref{eq5}). 
Equivalently, we obtain an unconstrained optimization problem by centering the basis functions. Let $b_{i,j}(x_{i,t})$ be replaced by $b_{i,j}(x_{i,t}) - \frac{1}{T}\sum_{t=1}^T b_{i,j}(x_{i,t})$. 
%
%Y_t \leftarrow Y_t - \frac{1}{T} \sum\limits_{t=1}^T Y_t ,
%\quad
%$$
%b_{i,j}(x_{t-i}) \leftarrow b_{i,j}(x_{t-i}) - \frac{1}{T} b_{i,j}(x_{t-i}).
%$$
By proper rearrangement, (\ref{eq3}) can be rewritten into a linear regression form
\begin{align} \label{eq4}
  \hat{e}_T =  \sum_{t=1}^T w_{T,t} (Y_{T} - \bm Z_T \bm \beta_T )^2 
\end{align}
where $\bm \beta_T$ is a $(1+\tilde{D} v) \times 1$ column vector to be estimated and $\bm Z_T$ is $1 \times (1+\tilde{D} v)$ row vector $\bm Z_T = [1,b_{1,1}(x_{1,T}),\ldots,b_{1,v}(x_{1,T}),\ldots,b_{\tilde{D},1}(x_{\tilde{D},T}),\ldots,b_{\tilde{D},v}(x_{\tilde{D},T})]$. 
Let $Z_T$ be the design matrix of stacking the row vectors $\bm Z_t, t= 1,\ldots, T$.  
Note that we have used $\bm \beta_T$ instead of a fixed $\bm \beta$ to emphasize that $\bm \beta_T$ may vary with time.
We have used bold style for vectors to distinguish them from matrices.
Let $W_T$ be the diagonal matrix whose elements are $w_{T,t},t=1,\ldots,T$. 
Then the optimal $\beta_T$ in (\ref{eq4}) can be recognized as the MLE of the following linear Gaussian model
\begin{align}
\bm Y_{T} = Z_T \bm{\beta}_T + \bm \v \label{eq6}
\end{align}
where $\bm \v \in \mathcal{N}(0,W_T^{-1})$.
Here, we have used $\mathcal{N} (\mu, \sigma^2)$ %, \Beta (a,b), \chi_k^2$ respectively 
to denote Gaussian distribution with mean $\mu$ and variance~$\sigma^2$.

To obtain a sharp model from large $L$, we further assume that the expansion of $f(\cdot,  \cdots, \cdot)$ is sparse, i.e., only a few additive components  $f_i$ are active. Clearly, selecting a sparse model is critical as models of large dimensions lead to inflated variance whereas models of small dimension lead to the lack-of-fit bias. 
To this end, we give independent Laplace priors for each sub-vector of $\bm{\beta}_T$ corresponding to each $f_i$. 
Our objective now reduces to obtaining the maximum a posteriori estimator (MAP)
\begin{equation}
\log p(\bm Y_T \mid \bm{\beta}_T) - \lambda_T \sum\limits_{i=1}^{\tilde{D}} \norm{\bm{\beta}_{T,i}}_2 . \label{eq13}
\end{equation}
%
%\begin{remark}
The above prior corresponds to the so called group LASSO.
The bold $\bm{\beta}_{T,i}$ is to emphasize that it is not a scalar element of $\bm{\beta}_{T}$ but a sub-vector of it. 
It will be interesting  to consider adaptive group LASSO \cite{zou2006adaptive}, i.e., to use $\lambda_{T,i}$ instead of a unified $\lambda_T$ and this is currently being investigated. We refer to \cite{huang2010variable} for a study of adaptive group LASSO for  batch estimation.  
%\end{remark}

\subsection{Implementation of SLANTS} \label{subsec:algo}

In order to solve the optimization problem given by (\ref{eq13}), we  build on an EM-based solution originally proposed for wavelet image restoration \cite{figueiredo2003algorithm}. This was further applied to online adaptive filtering for sparse linear models  \cite{babadi2010sparls}  and nonlinear models approximated by Volterra series \cite{mileounis2010adaptive,mileounis2010adaptive2}. 
The basic idea is to decompose the optimization (\ref{eq13}) into two parts that are easier to solve and iterate between them. One part involves linear updates, and the other involves group LASSO in the form of orthogonal covariance which leads to closed-form solution.

For now, we  assume that  the knot sequence
$t_{i,1},\ldots,t_{i,v}$ for each $i$ and  $v$ is fixed. 
Suppose that all the tuning parameters are well-defined. %We defer the discussion of sequentially updating the tuning parameters in section \ref{subsec:tuningParams}. 
We introduce an auxiliary variable $\tau_T$ that we refer to as the innovation parameter. This helps us to decompose the problem so that underlying coefficients can be iteratively updated. It also allows the sufficient statistics to be rapidly updated in a sequential manner.
The model in (\ref{eq6}) now can be rewritten as
\begin{align*}
\bm Y_{T} =  Z_T \bm \theta_T + W^{-\frac{1}{2}} \bm \v_1 , \ %\label{eq7} \\
\bm \theta_T = \bm \beta_T + \tau_T \bm  \v_2, 
\end{align*}
where
\begin{align} \label{eq10}
\bm \v_1 \in \mathcal{N}(0,I-\tau_T^2 W^{\frac{1}{2}}  Z_T  Z_T^\T W^{\frac{1}{2}} ) , 
\quad
\bm  \v_2 \in \mathcal{N}(0,I)
\end{align} 
%The estimation is usually done offline by, for example, gradient descent algorithm. In order to make it online, we rewrite the linear model as 
We treat $\bm \theta_T$ as the missing data,  so that an EM algorithm can be derived. 
By basic calculations similar to that of \cite{figueiredo2003algorithm}, we obtain the $k$th step of EM algorithm

{\it E step: }
\begin{align} \label{eq:Estep}
Q(\bm \beta \mid  \hat{\bm \beta}^{(k)}_T)=
-\frac{1}{2\tau_T^2} \norm{\bm \beta - \bm r^{(k)} }_2^2 - \lambda_T \sum\limits_{i=1}^{\tilde{D}} \norm{\bm \beta_i}_2 ,
\end{align}
where 
\begin{align} \label{eq:r}
\bm r^{(k)} = (I - \tau^2 A_T)  \hat{\bm \beta}^{(k)}_T
+ \tau^2 B_T ,
\\
A_T =  Z_T^\T W_T Z_T, \ 
\bm B_T = Z_T^\T W_T \bm Y_T .
\end{align}

{\it M step: }
$\hat{\bm \beta}^{(k+1)}_T$ is the maximum of   
$Q(\bm \beta \mid  \hat{\bm \beta}^{(k)}_T)$ given by 
\begin{align} \label{eq:Mstep}
\hat{\bm \beta}^{(k+1)}_{T,i} = \biggl[ 1-\frac{\lambda_T \tau_T^2}{\norm{\bm r^{(k)}_i}_2} \biggr]_{+} \bm r^{(k)}_{i}, \quad i=1,\ldots,\tilde{D}.
\end{align}

Suppose that we have obtained the estimator $\hat{\bm \beta}_T$ at time step $T$.
Consider the arrival of the $(T+1)$th point $(y_{T+1}, \bm x_{T+1})$,
respectively corresponding to the response and covariates (as a row vector) of time step $T+1$.
We first compute $\bm r^{(0)}_{T+1}$, the initial value of $\bm r$ to be input the EM at time step $T+1$:
$$
\bm r^{(0)}_{T+1} = (I - \tau^2 A_{T+1})  \hat{\bm \beta}_T
+ \tau^2 \bm B_{T+1},
$$
where 
\begin{align*}
A_{T+1} 
&= (1-\gamma_{T+1}) A_{T} + \gamma_{T+1}  \bm x_{T+1}^\T \bm x_{T+1},
\\
\bm B_{T+1}
&= (1-\gamma_{T+1}) \bm B_{T}+ \gamma_{T+1}  y_{T+1} \bm x_{T+1}^\T .
\end{align*}
Then we run the above EM for $K >0$ iterations to obtain an updated $ \hat{\bm \beta}_{T+1}$.

In the following Theorem, we show EM converges exponentially fast to the MAP of (\ref{eq13}). %Therefore, without mentioning, we assume EM has converged to the MAP and base further analysis on it. 
The proof is given in the appendix.

\vspace{0.1cm}

\begin{theorem}\label{thm:contraction}
 At each iteration, the mapping from $\hat{\bm \beta}_{T}^{(k)}$ to $\hat{\bm \beta}_{T}^{(k+1)}$ is a contraction mapping for any $\tau_T$, whenever the absolute values of all eigenvalues of $I - \tau^2 A_{T+1}$ are less than one. In addition, the error $\norm{ \hat{\bm \beta}_{T}^{(k+1)} - \hat{\bm \beta}_{T}}_2$ decays exponentially in $k$, where $\hat{\bm \beta}_{T}$ denotes the global minimum point of (\ref{eq13}). 
\end{theorem}

\vspace{0.1cm}

SLANTS can be efficiently implemented. In fact, by straightforward computations, the complexity of SLANTS at each time $t$ is $\Theta(\tilde{D}^2)$. 
Coordinate descent \cite{friedman2010regularization} is perhaps the most widely used algorithm for batch LASSO. 
Adapting coordinate descent to sequential setting has the same complexity for updating sufficient statistics. However, it does not have any convergence rate guarantees that we know of.
In contrast, SLANTS convergence is guaranteed by the above theorem.

%For standard Lasso (no group structure), coordinate descent \cite{friedman2010regularization} is widely used. For group Lasso, there are three R packages namely grplasso, grpreg, gglasso. grplasso implements the most widely used algorithm block coordinate descent or block coordinate gradient descent \cite{meier2008group}. If I understand correctly, they have to do a matrix transformation to orthogonalize the design matrix. This maybe computationally expensive and not suitable in sequential update. grpreg \cite{breheny2015group} uses block descent algorithm which is very similar to Meier's algorithm. They also require and advocate orthogonalization before applying the group penalty \cite{simon2012standardization}. gglasso \cite{yang2015fast} uses groupwise-majorization-descent (GMD) and we see the updating formula (algorithm 1) is very similar to SLANTS. But we claim from EM decomposition, $\gamma_{k}$ need not be the largest eigenvalue of $H^{(k)}$ in their paper, but can be anything satisfing that the EM decomposition is well defined. Therefore, we save computation of the eigenvalues in trade of the optimal shrinkage speed.

\subsection{The choice of tuning parameters: from a prequential perspective} 
 \label{subsec:tuningParams}

To evaluate the predictive power of an inferential model estimated from all the currently available data, ideally we would apply it to independent and identically generated datasets. However, it is not realistic to apply this cross-validation idea to real-world time series data,  since real data is not permutable and has a ``once in a lifetime'' nature. 
As an alternative, we adopt a prequential perspective \cite{dawid1984present} %:
that the goodness of a sequential predictive model shall be assessed by its forecasting ability.
%{%\color{blue} 
%\textit{
%[page 5] ``...We may wish to assess the overall adequacy of $P$ (prequential forecasting system) as a probabilistic explanation for $x$ (realized outcomes). It seems desirable that any such assessment should depend on P only through the sequence $\bm P = \{P_t\}$ (prequential forecast distributions produced by $P$) of forecasts that it in fact made. This requirement we shall call the Prequential Principle. It has an obvious analogy with the Likelihood Principle...''
%}
%}

Specifically, we evaluate the model in terms of the one-step prediction errors upon each newly arrived data point and subsequently tune the necessary control parameters, including regularization parameter $\lambda_t$ and innovation parameter $\tau_t$ (see details below).  Automatic tuning of the control parameters are almost a necessity in many real-world applications in which any theoretical guidance (e.g., our Theorem 2) may be insufficient or unrealistic. 
Throughout our algorithmic design, we have adhered to the prequential  principle and implemented the following strategies.

\textit{\bf The choice of $w_{T,t}$: }
$w_{T,t}$ is determined by
\begin{align*}
w_{1,1} &= \gamma_1, \quad
w_{t,j} = w_{t-1,j} (1-\gamma_t), \quad
w_{t,t} = \gamma_t, \, j=1,\ldots,t-1,
\end{align*}
and $\{\gamma_t\}$ is a nonnegative sequence which we refer to as the step sizes.
%\begin{remark}
It includes two special cases that have been commonly used in the literature. The first case is $\gamma_t=1/t$. It is easy to verify that $w_{T,t}=1/T , t=1,\ldots,T$  for any $T$. 
  This leads to the usual least squares. 
  The second case is $\gamma_t=c$ where $c$ is a positive constant. It gives $w_{T,t}=c(1-c)^{T-t},t=1,\ldots,T$. From (\ref{eq1}), the estimator of $f$ remains unchanged by rescaling $w_{T,t}$ by $1/c$, i.e. $w_{T,t}=(1-c)^{T-t}$ which is a series of powers of $1-c$. The value $1-c$ has been called the ``forgetting factor'' in the signal processing literature and used to achieve adaptive filtering \cite{babadi2010sparls}.

\textit{\bf The choice of $\tau_T$: }   
Because the optimization problem 
\begin{equation}
\log p(\bm Y_T \mid \bm  \beta_T) - \lambda_T \sum\limits_{i=1}^L \norm{\bm \beta_{T,i}}_2
\end{equation}
 is convex, as long as $\tau_T$ is proper, the EM algorithm converges to the global optimum regardless of what $\tau_T$ is.
But $\tau_T$ affects the speed of convergence of EM as $\lambda_T\tau_T^2$ determines how fast $\beta_T$ shrinks. Intuitively the larger  $\tau_T$ is, the faster is the convergence. Therefore we prefer $\tau_T$ to be large and proper.
A necessary condition for $\tau_T$ to be proper is to ensure that the covariance matrix of $\bm \v_1$ in 
\begin{align} 
\bm \v_1 \in \mathcal{N}(0,I-\tau_T^2 W^{\frac{1}{2}}  Z_T  Z_T^\T W^{\frac{1}{2}} ) , 
\quad
\bm  \v_2 \in \mathcal{N}(0,I)
\end{align} 
 is positive definite. Therefore, there is an upper bound $\bar{\tau}_T$ for $\tau_T$, and $\bar{\tau}_T$ converges to a positive constant $\bar{\tau}$ under some mild assumptions (e.g. the stochastic process $X_t$ is stationary). 
Extensive experiments have shown that $\bar{\tau}_T/2$ produces satisfying results in terms of model fitting. 
However,  it is not computationally efficient to calculate $\bar{\tau}_T$ at each $T$ in SLANTS. Nevertheless without computing $\bar{\tau}_T$, we can determine if $\tau_T < \bar{\tau}_T$ by checking the EM convergence. If $\tau_T$ exceeds $\bar{\tau}_T$, the EM would diverge and coefficients go to infinity exponentially fast. This can be proved via a similar argument to that of proof of Theorem 1. This motivates a lazy update of $\tau_T$ with shrinkage only  if EM starts to diverge. 
%In detail, we do the following:
%\begin{enumerate}
%\item Initialize $\tau_1$ by compute $\bar{\tau}_1$ and stay unchanged until either condition (b) or (c) satisfies.
%\item Shrink $\tau_T$ when EM diverges by recomputing $\bar{\tau}_T$ and redo the updates of coefficients.
%\item If $\tau_T$ does not change over some time $\Delta_{\tau}$, we enlarge $\tau_T$ by a predetermined value say $20\%$, to boost EM convergence rate. 
%\end{enumerate}

\textit{\bf The choice of $\lambda_T$: }  
On the choice of regularization parameter $\lambda_T$, different methods have been proposed in the literature. 
The common way is to estimate the batch data for a range of different $\lambda_T$'s, and select the one with minimum cross-validation error.  
To reduce the underlying massive computation required for such an approach, in the context of Bayesian LASSO  \cite{park2008bayesian}, \cite{bornn2010efficient} proposed an sequential Monte Carlo (SMC) based strategy to efficiently implement cross-validation.
The main proposal is to treat the posterior distributions educed by an ordered sequence of $\lambda_T$ as $\pi_t, t=0,1,\ldots$, the target distributions in SMC, and thus avoid the massive computation of applying Markov chain Monte Carlo (MCMC)  for each $\lambda$ independently. 
Another method is to estimate the hyper-parameter $\lambda_T$ via empirical Bayes method \cite{park2008bayesian}.
In our context, however, it is not clear whether the Bayesian setting with MCMC strategy can be efficient, as the dimension $Lv$ can be very large.  %A straightforward way is to use the same technique for updating $\tau_T$.
A possible implementation technique is to run three channels of our sequential modeling, corresponding to $\lambda_T^{-} = \lambda_T/\delta, \lambda_T, \lambda_T^{+}=\lambda_T*\delta$, where $\delta > 1$ is a small step size. The one with minimum average prediction error over the latest window of data was chosen as the new $\lambda_T$. For example, if $\lambda_T^{-} $ gives better performance, let the three channels be $\lambda_T^{-} / \delta, \lambda_T^{-}, \lambda_T^{-} * \delta$. We also have a parameter $\nu$ that favors bigger $\lambda$ when comparing prediction error because we have measurement uncertainty of prediction error. That is, the prediction error is multiplied by $(\nu^2, \nu, 1)$ in 3 channels. $\nu$ affects the model performance in our experiments, especially the model sparsity. We now specify $\nu$ heuristically.  It is ongoing work to fully understand how to specify $\nu$.
If there is an underlying optimal $\lambda^{\ast}$ which does not depend on $T$, we would like our channels to converge to the optimal $\lambda^{\ast}$ by gradually shrinking the stepsize $\delta$. Specifically in case that the forgetting factor $\gamma_t=1/t$, we let $\delta_T = 1+\frac{1}{T}(\delta-1)$ so that the step size $\delta_T \rightarrow  1$ at the same speed as weight of new data. 
%In detail, we adjust $\lambda_T$ in the following way:
%\begin{itemize}
%\item specify $\delta$ and initialize $\lambda_1$ as well as the three channels.
%\item Every other $\Delta_{\lambda}$ time, determine the latest average prediction error. If the middle channel is best, shrink the stepsize $\delta_T$ and change the left and right channel. Otherwise, left shift or right shift the channels as described above. In all cases, coefficients are carried on to new channels.
%\item A special situation is all channels produce 0 coefficients, which we call NULL model. To prevent stuck at NULL model, we immediately shrink all three channels and recompute the coefficients.
%\end{itemize}

\textit{\bf The choice of knots: }  
%While these polynomials may be of varying degrees, they are all recorded as polynomials of the same order k
The main difficulty in applying spline approximation is in determining the number of the knots to use and where they should be placed. 
Jupp \cite{jupp1978approximation} has shown that the data can be fit better with splines if the knots are free variables.
de Boor suggests the spacing between knots is decreased in proportion to the curvature (second derivative) of the data.
It has been shown that for a wide class of stationary process, the number of knots should be of the order of $O(T^{\zeta})$ for available sample size $T$ and some positive constant $\zeta$ to achieve a satisfying rate of convergence of the estimated nonlinear function to the underlying truth (if it exists) \cite{huang2004identification}.
Nevertheless, under some assumptions, we will show in Theorem~\ref{thm:step1} 
 that the prediction error can be upper bounded by an arbitrarily small number (which depends on the specified number of knots). It is therefore possible to identify the correct nonzero additive components in the sequential setting. 
%we use a fixed number of knots but allow the locations $t_{i,1},\ldots,t_{i,v+u}$ to be updated given new data. The derivation can be based upon the derivatives of $b_{i,j}(x) $ w.r.t. the knots \cite[Lemma 3.1]{jupp1978approximation}. 
%{\bf {\large This Sentence Needs to be fixed--I do not know what it says}}: 
On the other hand, using a fixed number of knots is  computationally desirable because sharp selection of significant spline basis/support in a potentially varying environment is computationally intensive.  
%To some extent it is similar to the idea of lasso for high dimensional covariates. 
%The updates of the knot locations are still under development.
%In our implementation, the knots are equally placed between the 5\% and 95\% sample quantiles of the data.  
It has been observed in our synthetic data experiments that the variable selection results are not very sensitive to the number of knots as long as this number is moderately large. % (e.g. around 10).
\section{Theoretical results} \label{sec:theory}
Consider the harmonic step size $\gamma_t = 1/t$. 
For now assume that the sequential update at each time $t$ produces $\hat{\bm \beta}_t$ that is the same as the penalized least squares estimator given batch data. 
We are interested in two questions. 
%\begin{enumerate}[1.]
%\item 
First, how to extend the current algorithm in order to take into account an ever-increasing number of dimensions? 
%\item 
Second, is it possible to select the ``correct'' nonzero components as sample size increases?   
%\end{enumerate}

The first question is important in practice as any prescribed finite number of dimensions/time series may not contain  the data-generating process, and it is natural to consider more candidates whenever more samples are obtained.
It is directly related to the widely studied high-dimensional regression for batch data. In the second question, we are not only interested in optimizing the prediction error but also to obtain a consistent selection of the true nonzero components. 
Moreover, in order to maintain low complexity of the algorithm, we aim to achieve the above goal by using a fixed number of spline basis. 
We thus consider the following setup. 
Recall the predictive Model (\ref{model:pred}) and its alternative form (\ref{model:00}). 
%\begin{align*}
%  X_{1,T} 
%  &
%  = f(X_{T-1},\ldots,X_{T-L})+\v_T \\
%  &
%  = f_{1,1}(X_{1,T-1})+\cdots+f_{1,L}(X_{1,T-L})+\cdots+
%  f_{D,1}(X_{D,T-1})+\cdots+f_{D,L}(X_{D,T-L})+\v_T
%. 
%\end{align*} 
We assume that $L$ is fixed while $D$ is increasing with sample size $T$ at certain rate. 
%Even when $L$ is fixed, the total number of regressors $\tilde{D}=DL$ is still increasing with sample size $T$.
%With a slight abuse of notation, we rewrite the above model as 
%%\begin{align} \label{model:reg}
%$  Y
%  %&
%  = f_{1}(X_1)+\cdots+f_{\tilde{D}}(X_{\tilde{D}})+\v_T
%$ %\end{align}
%with observations $Y=\{X_{1,t}\}$ and $[X_1,\ldots,X_{\tilde{D}}]$ $=$ $\{[X_{1,t-1},\ldots,X_{1,t-L}, \ldots, X_{D,t-1} , \ldots, X_{D,t-L}]\}$, where $\tilde{D} = DL$. 

Following the setup of \cite{stone1985additive}, we suppose that each $X_d$ takes values from a compact interval $[a,b]$. Let $[a,b]$ be partitioned into $J$ equal-sized intervals $\{I_{j}\}_{j=1}^J$, and let $\mathfrak{F}$ denote the space of polynomial splines of degree $\ell \geq 1$ consisting of functions $g(\cdot)$ satisfying 1) the restriction of $g(\cdot)$ to each interval is a polynomial of degree $\ell$, and 2) $g(\cdot) \in C^{\ell-1}[a,b]$ ($\ell-1$ times continuously differentiable). Typically, splines are called linear, quadratic or cubic splines accordingly as $\ell=1,2$, or $3$.
There exists a normalized B-spline basis $\{b_{j}\}_{j=1}^{v}$ for $\mathfrak{F}$, where $v = J+\ell$, and any $f_i(x) \in \mathfrak{F}$ can be written in the form of (\ref{eq:spline}).
Let $k$ be a nonnegative integer,  $\beta \in (0,1]$ that $p=k+\beta>0.5$, and $M>0$. Suppose each considered (non)linear function $f$ has $k$th derivative, $f^{(k)}$, and satisfies the Holder condition with exponent $\beta$:
$|f^{(k)}(x)-f^{(k)}(x') | < M |x-x'|^{\beta}
$
for $x,x' \in [a,b]$. Define the norm $\norm{f}_2 = \sqrt{\int_a^b f(x)^2 dx}$. 
Let $f^{*} \in \mathfrak{F}$ be the best $L_2$ spline approximation of $f$.  
Standard results on splines imply that $\norm{f_d-f^{*}_d}_{\infty} = O(v^{-p})$ for each $d$. The spline approximation is usually an estimation under a mis-specified model class (unless the data-generating function is low-degree polynomials), and   large $v$ narrows the distance to the true model. We will show that for large enough $v$, it is possible to achieve the aforementioned two goals. 
To make the problem concrete, we need the following assumptions on the data-generating procedure.
\begin{assumption} \label{ass:A1}
The number of additive components is finite and will be included into the candidate set in finite time steps.
In other words, 
there exists  a ``significant'' variable set $S_0 = \{i_1,\ldots,i_{D_0}\}$ such that  1) $f_{d}(x) \neq 0$ for each $d \in S_0$, 2) $f_d(x) \equiv 0$ for $d \notin S_0$, and 3) both $D_0$ and $i_{D_0}$ are finite integers that do not depend on sample size $T$.   
\end{assumption}

We propose two steps for a practitioner targeting two goals given below. 

%\begin{enumerate}[Step 1.]
\textit{\bf Step 1. (unbiasedness)}
%\item 
 This step aims to discover the significant variable set with probability close to one as more data is collected.
  The approach is to minimize the objective function in (\ref{eq13}), and it can be efficiently implemented using the proposed sequential algorithm in Section~\ref{subsec:algo} with negligible error (Theorem~\ref{thm:contraction}).  In the case of equal weights $w_{T,t} = 1/T$, it can be rewritten as 
\begin{equation} \label{eq:grouplasso}
 \norm{Y_T - Z_T \bm  \beta_T}_2^2 + \tilde{\lambda}_T \sum\limits_{i=1}^{\tilde{D}} \norm{\bm \beta_{T,i}}_2  
\end{equation}
where $\tilde{\lambda}_T = 2 T \lambda_T$.
Due to Assumption~\ref{ass:A1},  the significant variable set $S_0$ is included in the candidate set $\{1,\ldots,\tilde{D}\}$ for sufficiently large $T$. 
Our selected variables are those whose group coefficients are nonzero, i.e. $S_1 = \{d: 1 \leq d \leq \tilde{D}, \hat{\bm \beta}_{T,d} \neq \bm 0\}$. We are going to prove that all the significant variables will be selected by minimizing (\ref{eq:grouplasso}) with appropriately chosen $\tilde{\lambda}_T $, i.e., $S_0 \subseteq S_1$. 

\textit{\bf Step 2. (minimal variance) }
%\item  (minimal variance) 
The second step is optional and it is applied only when a practitioner's goal is to avoid selecting any redundant variables outside  $S_0$. 
To achieve consistency of variable selection, we use the variables with nonzero estimated coefficients from Step 1 as candidate variables. Then we apply a BIC-type penalized method on future data points to further remove redundant variables. 
Suppose that we obtain a candidate set $S_1$ of $\tilde{D}$ variables (where $S_0 \subseteq S_1$ from the Step 1) from fitting the sequential data upto $T_1=T/2$ and we perform the further procedure on data from $(T_1+1)$ to $T$.   
    Since a thorough search over all subsets of variables is computationally demanding, we use a backward stepwise procedure. We start with the set of $S_1 $ selected variables, delete one variable at a time by  minimizing the MSE of a spline model with $v_T = T_1^{\zeta}$ number of equally spaced knots. Specifically, suppose that at step $k$ ($k=1,2,\ldots$), the survived candidate models are indexed by $ \mathcal{S}^{(k)}$, we solve the least-squares problem for each $\bar{d} \in \mathcal{S}^{(k)}$
    \begin{align} \label{eq:BIC}
  \hat{e}^{(k)}_{\bar{d}} = \min_{\mu,c_{d,j}} \sum_{t=T_1+1}^T \biggl(Y_{t} - \mu - \sum_{d \in \mathfrak{S} } \sum_{j=1}^{v_T} c_{d,j} b_{d,j}(X_{d,t}) \biggr)^2 ,
  \end{align}
  where $\mathfrak{S} = \mathcal{S}^{(k-1)}- \{\bar{d} \}$
  and select $\bar{d} = \bar{d}_k^{*}$ that minimize the $\hat{e}^{(k)}_{\bar{d}}$ with minimum denoted by $\hat{e}^{(k)}$. Let  $\mathcal{S}^{(k)} = \mathcal{S}^{(k-1)} - \bar{d}_k^{*}$.
  By default, we let $\mathcal{S}^{(0)} = S_1$ and use $\hat{e}^{(0)}$ to denote the minimum of (\ref{eq:BIC}) with $\mathfrak{S} = S_1$.
  If $\hat{e}^{(k-1)} - \hat{e}^{(k)} < (v_T \log T_1 )/ T_1$ (the gain of goodness of fit is less than the incremented BIC penalty), then we stop the procedure and output $S_2 = \mathcal{S}^{(k-1)}$; otherwise we proceed to the $(k+1)$th iteration.      
    We prove that the finally selected subset $S_2$ satisfies $\lim_{T \rightarrow \infty}\P(S_2 = S_0) = 1$. % as $T \rightarrow \infty$.
%\end{enumerate}

Before we proceed to the theoretical result, we introduce some necessary assumptions and their interpretations. 

\begin{assumption} \label{ass:A2}
  There is a positive constant $c_0$ such that $\min_{d \in S_0} \norm{f_d}_2 \geq c_0$. 
\end{assumption}

\begin{assumption} \label{ass:A3}
  The noises $\v_t$ are sub-Gaussian distributed, i.e., $E(e^{w\v_t}) \leq e^{w^2\sigma^2/2}$ for a constant $\sigma>0$ and any $w \in \mathbb{R}$.   
\end{assumption}

\begin{assumption} \label{ass:A4}
  Suppose that $S_1$ is a finite subset of $\{1,\ldots, \tilde{D}\}$. 
  In addition, the ``design matrix'' $Z_{S_1}$ satisfies $  Z_{S_1}^\T Z_{S_1} / T \geq \kappa$
  for a positive constant $\kappa$ that depend only on $v$ (the number of splines). 
\end{assumption}

We use $o_p(1)$ and $O_p(1)$ to denote a sequence of random variables that converges in probability to zero, and that is stochastically bounded, respectively. We use $O(1)$ to denote a bounded deterministic sequence. 

\vspace{0.1cm}

%theorem 
\begin{theorem} \label{thm:step1}
  Suppose that Assumptions~\ref{ass:A1}-\ref{ass:A4} hold. Then for any given $v$ it holds that 
  \begin{align} 
  \norm{\bm \beta_{\tilde{S}_1} - \hat{ \bm\beta}_{\tilde{S}_1} }_2^2
  \leq &8c_2 v^{-2p} /\kappa  + O_p(T^{-1} \log \tilde{D} ) +
  	O_p(T^{-1}) + O(T^{-2} \tilde{\lambda}^2) \label{bound}
 \end{align}
for some positive constant $c_2$. If we further assume that $\log \tilde{D} = o(T)$, $\tilde{\lambda} = o(T)$, then there exists a constant $c_1>0$ such that for all $v > c_1 c_0^{-1/p} \max\{ 1, c_0^{-\frac{1}{p(2p+1)}}\}$,  %then all the significant variables will be selected through the above step 1 with probability going to one as $T$ tends to infinity, i.e., 
$\lim_{T \rightarrow \infty} \P (S_0 \subseteq S_1) = 1$. 
\end{theorem}

\vspace{0.1cm}

\begin{remark} \label{remark:ass}
Theorem~\ref{thm:step1} gives an error bound between the estimated spline coefficients with the oracle, where the first term is dominating. As a result, if $v$ is sufficiently large, then it is guaranteed that $S_0$ will be selected with probability close to one. 
We note that the constant $c_1$ depends  only on the true nonlinear function and the selected spline basis function. 
In proving Theorem~\ref{thm:step1},  
Assumption~\ref{ass:A2}-\ref{ass:A3} serve as standard conditions to ensure that a significant variable is distinguishable, and that any tail probability could be well bounded. 
Assumption~\ref{ass:A4} is needed to guarantee that if the estimated coefficients $\hat{\beta}$ produces low prediction errors, then it is also close to the true (oracle) coefficients. This assumption is usually guaranteed by requiring $\tilde{\lambda} > c \sqrt{T \log D}$. See for example \cite{huang2010variable,hastie2015statistical}. 
\end{remark}

To prove the consistency in step 2, we also need the following  assumption (which further requires that the joint process is strictly stationary and strongly mixing). 
\begin{assumption} \label{ass:A6}
$\sup_x \{E(|Y_t|^r|\bm X_t = x) \} < \infty$ for some $r>2$.
\end{assumption}

The $\alpha$-mixing coefficient is defined as %TODO math mode
  $$
  \alpha_S(j) = \sup 
  \{ P(E_y \cap E_x)-P(E_y)P(E_x):
    E_y \in \sigma(\{(Y_{\tilde{t}},X_{d,\tilde{t}},d \in S):\tilde{t}\leq n\}), 
    E_x \in \sigma(\{(Y_{\tilde{t}},X_{d,\tilde{t}},d \in S):\tilde{t}\geq n+j\})
  \},
  $$
  where $\sigma(\cdot)$ denotes the $\sigma$-field generated by the random variables inside the parenthesis.
\begin{assumption} \label{ass:A5}
  The process $\{(X_{d,t},d \in S_1)\}$ is strictly stationary, and the joint process $\{(Y_t,X_{d,t},d \in S_1)\}$ is $\alpha$-mixing with coefficient 
  $$\alpha_{S_1}(j) \leq  \min\{O(j^{-2.5\zeta/(1-\zeta)}), O(j^{-2r/(r-2)}) \}.$$
\end{assumption}

\vspace{0.1cm}

\begin{theorem} \label{thm:step2}
  Suppose that Assumptions~\ref{ass:A1}-\ref{ass:A5} hold, then the $S_2$ produced by the above step 2 satisfies %asymptotically equals $S_0$, i.e., 
  $\lim_{T \rightarrow \infty} \P ( S_2 = S_0) = 1$.  
\end{theorem}

\vspace{0.1cm}

\section{Numerical results}
\label{sec:Num_Results}
% ======
In this section, we present experimental results to demonstrate the theoretical results and the advantages of SLANTS on both synthetic and real-world datasets. 
%Throughout the experiments, we implement our algorithm  based upon philosophies discussed in Subsection~\ref{subsec:tuningParams}% and the corresponding part in our supplementary material

%=============================================
\subsection{Synthetic data experiment}

We have carried out comprehensive experiments to show the performance of SLANTS in modeling nonlinear relation in cases where the data-generating model is fixed over time, is  varying over time, or has a large dimensionality. 
%This part has been included in the supplementary material. 

\subsubsection{Synthetic data experiment: modeling nonlinear relation in stationary environment} \label{subsubsec:exp1}

The purpose of this experiment is to show the performance of SLANTS in stationary environment where the data-generating model is fixed over time. 
We generated synthetic data using the following nonlinear model
\begin{align*}
X_{1,t}&=\epsilon_{1,t}, \quad
X_{2,t}=0.5X_{1,t-1}^2-0.8X_{1,t-7}+0.2\epsilon_{2,t}, \,t=1,\ldots,500,
\end{align*}
where $\epsilon_{1,t}$ and $\epsilon_{2,t}$ are i.i.d. Gaussian with mean zero and variance one. The initial $L$ values of $X_{2,t}$ are set to zero. The goal is to model/forecast the series $X_{2,t}$. We choose $L=8$, and place $v=10$ quadratic splines in each dimension. The knots are equally spaced between the $0.01$ and $0.99$ quantiles of observed data. We choose forgetting factor $\gamma_t=1/t$ to ensure the convergence. 

\begin{figure} [!h] 
        \vspace*{-0.0 in}
	\begin{center}
    \hspace{-0.0 in}\includegraphics[width=4.5in]{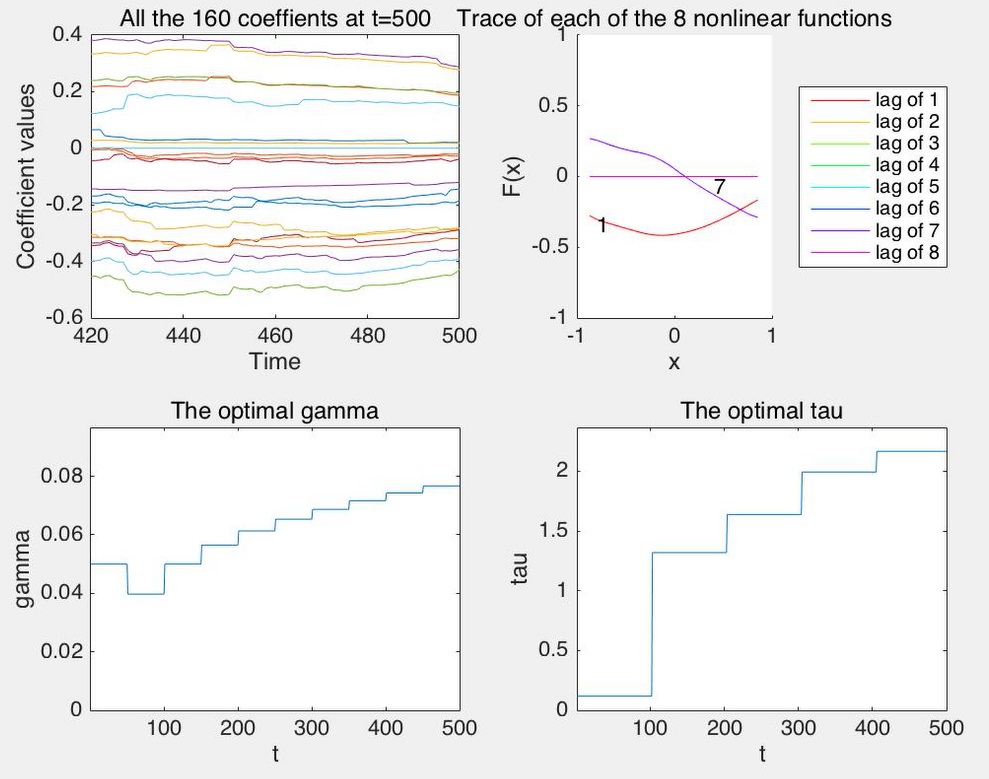}
    \end{center}
    \vspace*{-0.0in}
    \caption{Four subplots show the estimated coefficients of splines, nonlinear functions, and trace plots of automatically-tuned regularization parameter $\lambda_t$ and innovation parameter~$\tau_t$. A demo video is available at  {\color{blue} goo.gl/PJI2uJ }
}
    \label{fig:video1}
    \vspace*{-0.0in}
    \end{figure}

Simulation results are summarized in Fig \ref{fig:video1}. %, available from the following dropbox link: 
%{\color{blue} https://www.dropbox.com/s/k8c7zgjuuldxvbg/Video1.avi?dl=0 }
%
The left-top plot shows the convergence of all the $2 \times 8 \times 10=160$ spline coefficients. The right-top plot shows how the eight nonlinear components $f_d,d=1,\ldots,8$ evolve, where the number 1-8 indicate each additive component (splines). The values of each function are centralized to zero for identifiability. The remaining two plots show the optimal choice of control parameters $\lambda_t$ and $\tau_t$ that have been automatically tuned over time. In the experiment, the active components $f_1$ and $f_7$ are correctly selected and well estimated. It is remarkable that the convergence is mostly achieved after only a few incoming points (less than the number of coefficients 160).

%=============================================  
\subsubsection{Synthetic data experiment: modeling nonlinear relation in adaptive environment}

The purpose of this experiment is to show the performance of SLANTS in terms of prediction and nonlinearity identification when the underlying date generating model varies over time. 

\begin{figure}[bt]
        \vspace*{-0.0 in}
	\begin{center}
    \hspace{-0.0 in}\includegraphics[width=4.5in]{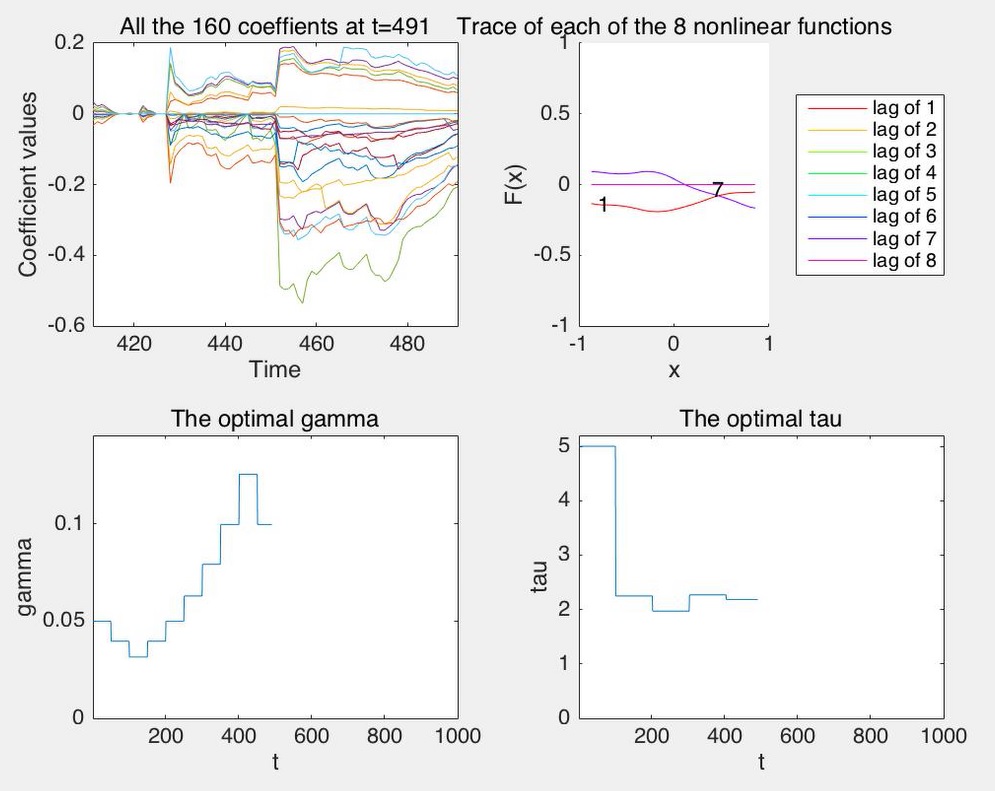}
    \\
    \hspace{-0.0 in}\includegraphics[width=4.5in]{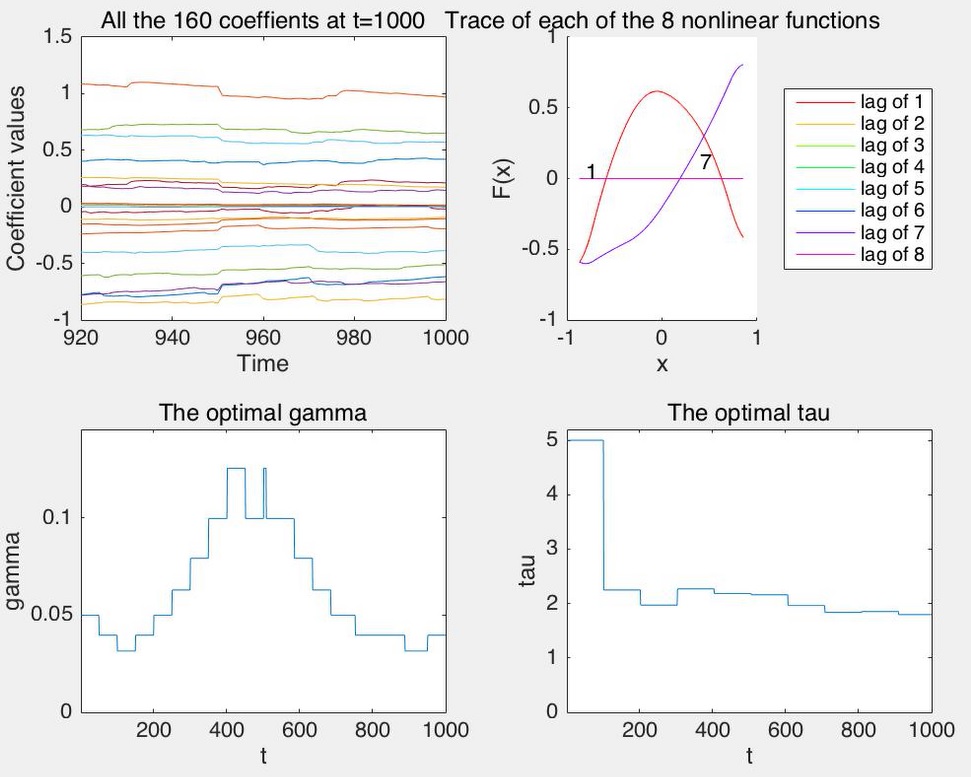}
    \end{center}
    \vspace*{-0.0in}
    \caption{Two plots stacked vertically each consists of four subplots showing the estimated coefficients of splines, nonlinear functions, and trace plots of automatically-tuned regularization parameter $\lambda_t$ and innovation parameter~$\tau_t$ at time $t=491$ and $t=1000$ respectively. 
    A demo video is available at {\color{blue} goo.gl/Vycrve }
    }
    \label{fig:video2}
    \vspace*{-0.0in}
    \end{figure}

We have generated a synthetic data using the following nonlinear model where there is a change at time $t=500$,
\begin{align*}
X_{1,t}&=\epsilon_{1,t}, \quad
X_{2,t}=0.5X_{1,t-1}^2-0.8X_{1,t-7}+0.2\epsilon_{2,t}, \,t=1,\ldots,500, \\
X_{1,t}&=u_{1,t}, \quad
X_{2,t}=-2X_{1,t-1}^2+\exp(X_{1,t-7})+0.2\epsilon_{2,t}, \,t=501,\ldots,1000,
\end{align*}
where $\epsilon_{1,t}$ and $\epsilon_{2,t}$ are i.i.d. Gaussian with mean zero and variance one. $u_{1,t}$ are i.i.d. uniform on $[-1,1]$. The initial $L$ values of $X_{2,t}$  are set to zero. The goal is to model the series $X_{2,t}$. Compared with the previous experiment, the only difference is that the forgetting factor is set to $\gamma=0.99$ in order to track potential changes in the underlying true model. Fig \ref{fig:video2} shows that the sequential learning algorithm successfully tracked a change after the change point $t=500$. The top plot in Fig \ref{fig:video2} shows the fit right before the change. It successfully recovers the quadratic curve of lag $1$ and linear effect of lag $7$. The bottom plot in Fig \ref{fig:video2} shows the fit at $t=1000$. It successfully finds the exponential curve of lag $7$ and reversed sign of the quadratic curve of lag $1$. From the bottom left subplot we can see how the autotuning regularization parameter decreases since the change point $t=500$. 
%
%The simulation results are summarized in Video 2, available from the following dropbox link: 
%{\color{blue} https://www.dropbox.com/s/bhxmckqexvvttcg/Video2.avi?dl=0 }

%=============================================
%=============================================  
\subsubsection{Synthetic data experiment: causal discovery for multi-dimensional time series}

The purpose of this experiment is to show the performance of SLANTS in identifying nonlinear functional relation (thus Granger-type of causality) among multi-dimensional time series. 

We have generated a 9-dimensional time series  using the following nonlinear network model, 
\begin{align*}
X_{1,t}&=\epsilon_{1,t}\\
X_{2,t}&=0.6X_{3,t-1}+\epsilon_{2,t}\\
X_{3,t}&=0.3X_{4,t-2}^2+\epsilon_{3,t}\\
X_{4,t}&=0.7X_{5,t-1}-0.2X_{5,t-2}+\epsilon_{4,t}\\
X_{5,t}&=-0.2X_{2,t-1}^2+\epsilon_{5,t}\\
X_{6,t}&=0.5X_{6,t-2}+1+\epsilon_{6,t}\\
X_{7,t}&=2\exp(-X_{7,t-2}^2)+\epsilon_{7,t}\\
X_{8,t}&=6X_{7,t-1}-5X_{9,t-2}+\epsilon_{8,t}\\
X_{9,t}&=-X_{6,t-1}+0.9X_{7,t-2}+\epsilon_{9,t}
\end{align*}
where $\epsilon_{1,t}$ and $\epsilon_{2,t}$ are i.i.d.
The initial $L$ values are set to zero. The goal is to model each dimension and draw sequential causality graph based on the estimation. We choose $L=2$, $M=10$ and $\gamma_t=1/t$. For illustration purpose, we only show the estimation for $X_{9,t}$. The left-top plot shows the 9 dimensional raw data that are sequentially obtained. The right-top plot shows the convergence of the $9\times 2\times 10=180$ coefficients in modeling $X_{9,t}$. The right-bottom plot shows how the nonlinear components $f: X_{6,t-1} \rightarrow X_{9,t}$ and  $f: X_{7,t-2} \rightarrow X_{9,t}$ evolve. Similar as before, the values of each function are centralized to zero for identifiability. The left-bottom plot shows the causality graph, which is the digraph with black directed edges and edge labels indicating functional relations. For example, in modeling $X_{9,t}$, if $F(X_{6,t-1})$ is nonzero, then draw a directed graph from 6 to 9 with edge label 1; if both $F(X_{6,t-1})$  and $F(X_{6,t-2})$ are nonzero, then draw a directed graph from 6 to 9 with edge label 12. The true causality graph (determined by the above data generating process) is draw together, in red thick edges. From the simulation, the discovered causality graph quickly gets close to the truth. %In fact, we can theoretically prove this convergence under mild conditions. The theoretical results will be included in a paper we are preparing.
% 
% The simulation results are summarized in Video 3, available from the following dropbox link:  
%{\color{blue} https://www.dropbox.com/s/zg1gsy52htdj8td/Video3.avi?dl=0 }
%  
  
    \begin{figure} [!h]
        \vspace*{-0.0 in}
	\begin{center}
    \hspace{-0.0 in}\includegraphics[width=4.5in]{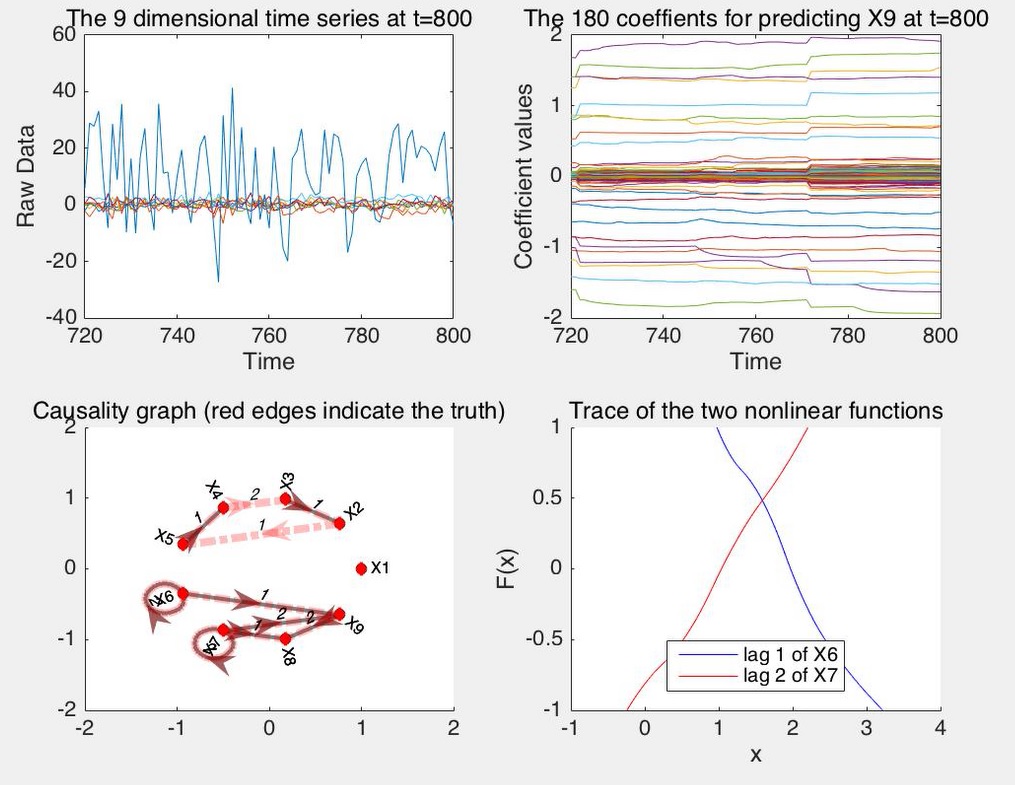}
    \end{center}
    \vspace*{-0.0in}
    \caption{Four subplots show the nine time-series data, convergence of the coefficients, causality graph, and trace plot of the nonlinear functions. A demo video is available at {\color{blue} goo.gl/sWM1oh }
    }
    \label{fig:video3}
    \vspace*{-0.0in}
    \end{figure}
    
\subsubsection{Synthetic data experiment: computation cost comparison}

The purpose of this experiment is to show SLANTS is computationally efficient by comparing it to standard batch group LASSO algorithm. We use the same data generating process in the first synthetic data experiment %\ref{subsubsec:exp1} 
but let the number of data points $T$ increases as $T=1000, 2000, 3000$. We compare to the R package 'grplasso' \cite{meier2008group} which implemented a widely used group LASSO algorithm. The result is shown in Table~\ref{tab:computation}. 
The table shows the time in seconds for SLANTS and grplasso to run through a dataset sequentially with different size $T$. Each run is repeated $3$ times and the standard error of running time is shown in parenthesis. 
From Table~\ref{tab:computation}, the computational cost of SLANTS grows linearly with $T$ while grplasso grows quadratically. %but scales poorer than grplasso with increasing $D$.
%as we explained in section \ref{subsec:algo}.

\begin{table}[h] 
\centering
\caption{The table shows the computational cost in seconds with standard error in parenthesis for the two methods with increasing $T$.}
\label{tab:computation}
\begin{tabular}{|c|c|c|c|}
\hline
         & T=1000  & T=2000  & T=3000 %& \multicolumn{2}{c|}{N=1000} 
         \\ \hline
         %& \multicolumn{3}{c|}{D = 2} &  D=4     & D=8     \\ \hline
SLANTS   & 24.29(0.12) & 53.09(3.80) & 81.36(3.42) %& 68.71(7.46)     & 149.68(0.51)    
\\ \hline
grplasso &  25.37(0.42) & 75.68(0.36) & 151.41(5.58) %& 50.64(9.80)   & 74.90(0.93)   
\\ \hline
\end{tabular}
\end{table}
    
%In summary, the algorithm is designed for the purpose of sequential estimation of a large dimensional time series with potentially nonlinear functional, which achieves a good balance between the goodness of fit and the model complexity.

%%=============================================  
%\subsection{Synthetic data experiment: enhancing predictive power by adapting to new environments}
%
%
%
%
%
%
%
%
%
%
%
%
%
%
%%=============================================
%%=============================================  
%\subsection{Synthetic data experiment: causal discovery for multi-dimensional time series}

%we generate a 9 dimensional data using the following nonlinear model, 
%The goal is to model each dimension and draw an online causality graph based on the estimation. 
%
%In summary, the algorithm is designed for the purpose of sequential estimation of a large dimensional time series with potentially nonlinear functional, which achieves a good balance between the goodness of fit and the model complexity. 
%

%=============================================
\subsection{Real data experiment: Boston weather data from 1980 to 1986}

\begin{figure} [tb]
        \vspace*{-0.0 in}
	\begin{center}
    \hspace{-0.0 in}
    \includegraphics[width=6in]{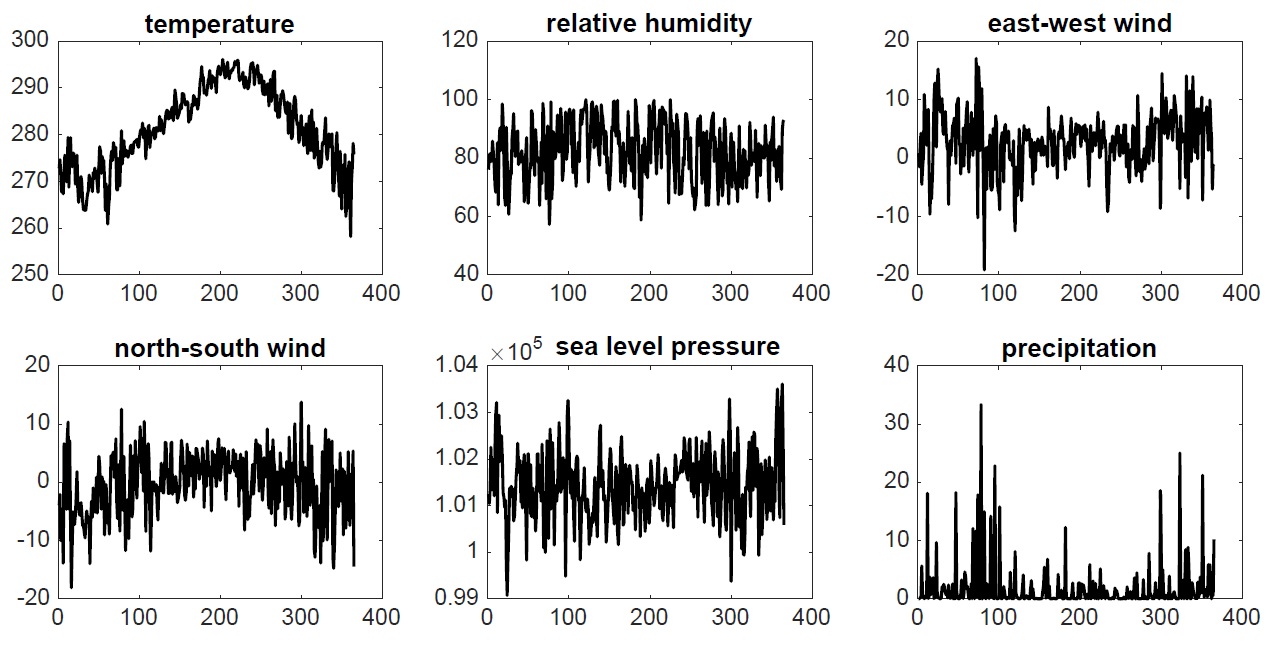}
    \end{center}
    \vspace*{-0.0in}
    \caption{A graph showing the raw data of (a) temperature (K), (b) relative humidity (\%), (c) east-west wind (m/s), (d) north-south wind (m/s), (e) sea level pressure (Pa), and (f) precipitation (mm/day).}
    \label{fig:envi_raw}
    \vspace*{-0.0in}
    \end{figure}

In this experiment, we study the daily Boston weather data from 1980 Jan to 1986 Dec.
%Boston daily weather data from 1980 Jan to 1986 Dec, daily data, 
with $T=2557$ points in total.
The data is a six-dimensional time series, with each dimension corresponding respectively to temperature (K), relative humidity (\%), east-west wind (m/s), north-south wind (m/s), sea level pressure (Pa), and precipitation (mm/day).
In other words, the raw data is in the form of $X_{d,t},d=1,\ldots,6,t=1,\ldots,T$. We plot the raw data from 1980 $X_{d,t}, d=1,\ldots,6, t= 1,\ldots,366$ in Fig.~\ref{fig:envi_raw}. 
%Boston daily weather data from 1980 Jan to 1986 Dec, daily data, 
%with $T=2557$ points in total.
%The data is a six-dimensional time series, with each dimension corresponding respectively to temperature (K), relative humidity (\%), east-west wind (m/s), north-south wind (m/s), sea level pressure (Pa), and precipitation (mm/day).
%In other words, the raw data is in the form of $X_{d,t},d=1,\ldots,6,t=1,\ldots,T$.

\begin{figure} [tb]
        \vspace*{-0.0 in}
	\begin{center}
    \hspace{-0.0 in}\includegraphics[width=6.5in]{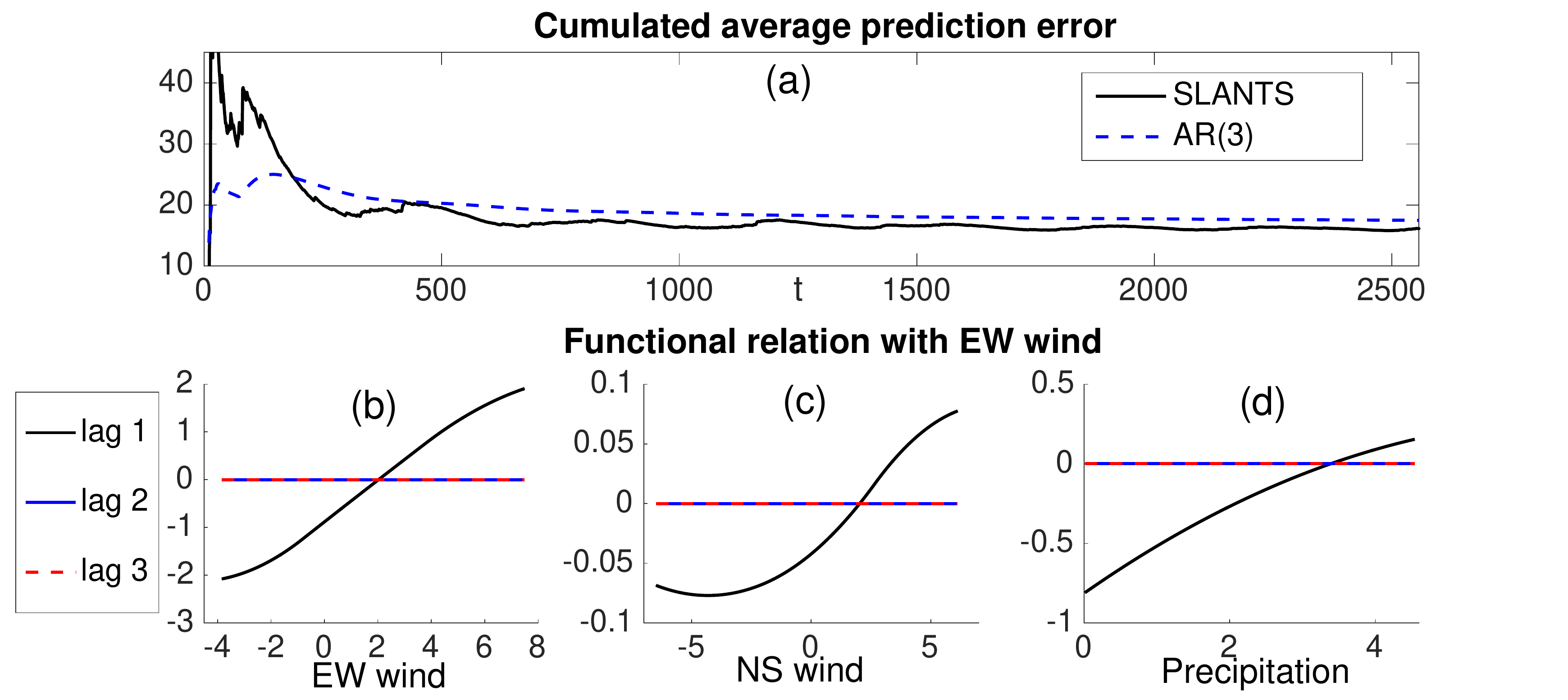}
    \end{center}
    \vspace*{-0.2in}
    \caption{A graph showing (a) the cumulated average one-step ahead prediction error produced by two approaches, and east-west wind (m/s) decomposed into nonlinear functions of lagged values of (b) east-west wind, (c) north-south wind (m/s), and (c) precipitation (mm/day). The functions were output from SLANTS at the last time step $t=T$.}
    \label{fig:envi_X3_cumAvePredErr_vsSequentialAR3}
    \vspace*{-0.2in}
    \end{figure}

We compare the predictive performance of SLANTS with that of a linear model. We chose the autoregressive model of order $3$ (denoted by AR$(3)$) as the representative linear model. The order was chosen by applying either the Akaike information criterion \cite{akaike1969fitting,akaike1998information} or the sample partial autocorrelations \cite{anderson1962determination} to the batch data of $T$ observations.  
We started processing the data from $t_0=10$, and for each $t=t_0+1,\ldots,T$ the one-step ahead prediction error $\hat{e}_{t} $ was made by applying AR$(3)$ and SLANTS to the currently available $t-1$ observations. The cumulated average prediction error at time step $t$ is computed to be $\sum_{t=t_0+1}^{t} \hat{e}_{t} / (t-t_0)$. Fig.~\ref{fig:envi_X3_cumAvePredErr_vsSequentialAR3}(a). 
At the last time step, the significant (nonzero) functional components are the third, fourth, and sixth dimension, corresponding to EW wind, NS wind, precipitation, have been plotted in Fig.~\ref{fig:envi_X3_cumAvePredErr_vsSequentialAR3}~(b), (c), (d), respectively.  
From the plot,
the marginal effect of $X_{4,t}$ on $X_{3,t+1}$ is clearly nonlinear. It seems that the correlation is low for $X_{4,t} < 0$ and high for $X_{4,t}>0$.
In fact, if we let $\mathfrak{T} = \{t: X_{4,t} > 0 \}$, the correlation of $\{ X_{4,t}: t \in \mathfrak{T}\}$ with $\{ X_{3,t+1}: t \in \mathfrak{T}\}$ is $0.25$ (with p value $1.4\times 10^{-8}$) 
while $\{ X_{4,t}: t \not\in \mathfrak{T}\}$ with $\{ X_{3,t+1}: t \not\in \mathfrak{T}\}$ is $-0.05$ (with p value $0.24$) 

%=============================================
\subsection{Real data experiment: the weekly unemployment data from 1996 to 2015}

\begin{figure} [tb]
        \vspace*{-0.0 in}
	\begin{center}
    \hspace{-0.0 in}\includegraphics[width=6in]{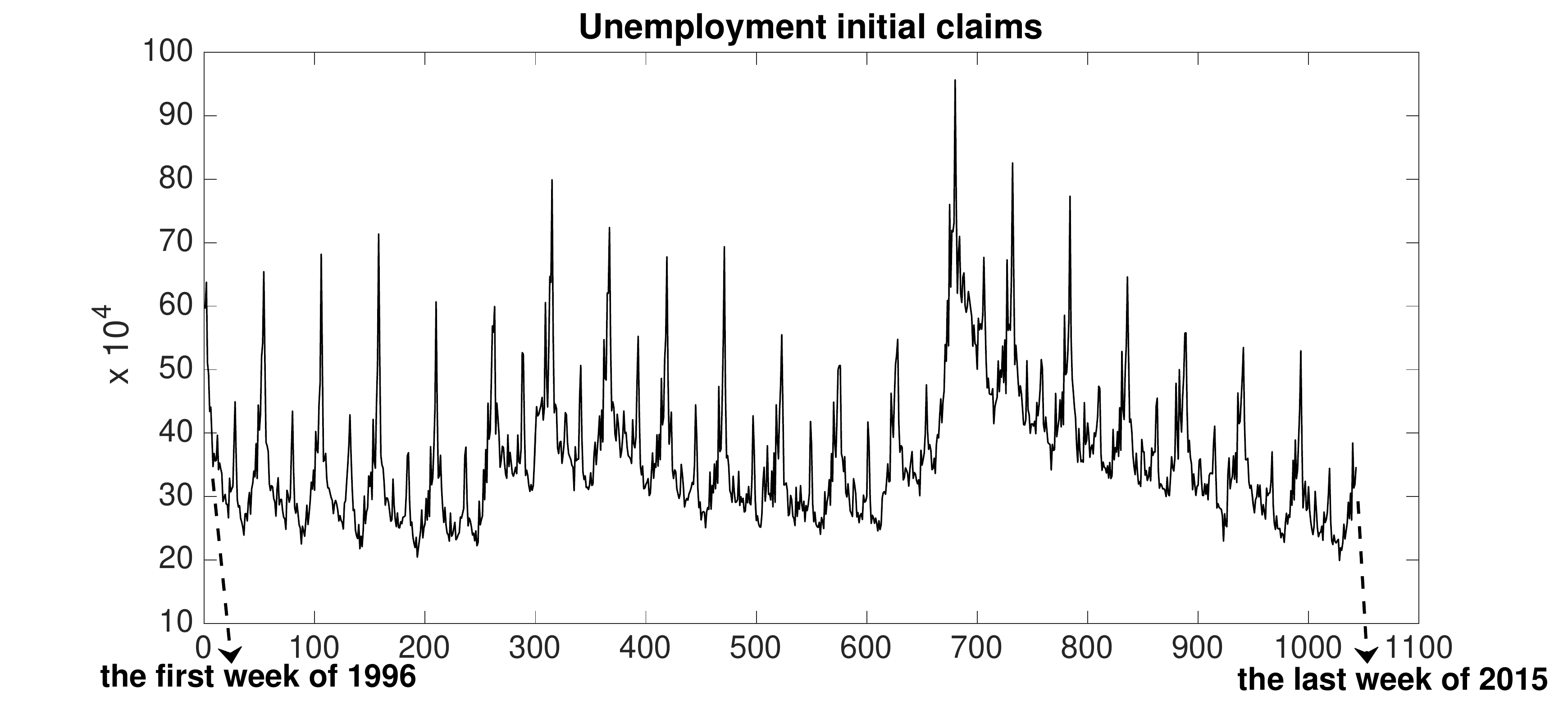}
    \end{center}
    \vspace*{-0.0in}
    \caption{A graph showing the raw data of the number of unemployment initial claims.}
    \label{fig:unemployment:data}
    \vspace*{-0.0in}
    \end{figure}

In this experiment, we study the US weekly unemployment initial claims from Jan 1996 to Dec 2015.
The data is a one-dimensional time series with $T=1043$ points in total. 
we plot the raw data in  Fig.~\ref{fig:unemployment:data}.

Though the data exhibits strong cyclic pattern, it may be difficult to perform cycle-trend decomposition in a sequential environment. We explore the power of SLANTS to do lag selection to compensate the lack of such tools.

We compare three models. The first model, AR(5), is linear autoregression with lag order 5. The lag order was chosen by the sample partial autocorrelations to the batch data. The second and third are SLANTS(1) with linear spline and SLANTS(2) with quadratic splines. SLANTS(1) have 1 spline per dimension, which is exactly LASSO with auto tuning penalty parameter in SLANTS. SLANTS(2) have 8 splines per dimension. We allow SLANTS to select from a maximum lag of 55, which is roughly the size of annual cycle of 52 weeks.

\begin{figure} [!h]
        \vspace*{-0.0 in}
	\begin{center}
    \hspace{-0.0 in}\includegraphics[width=6in]{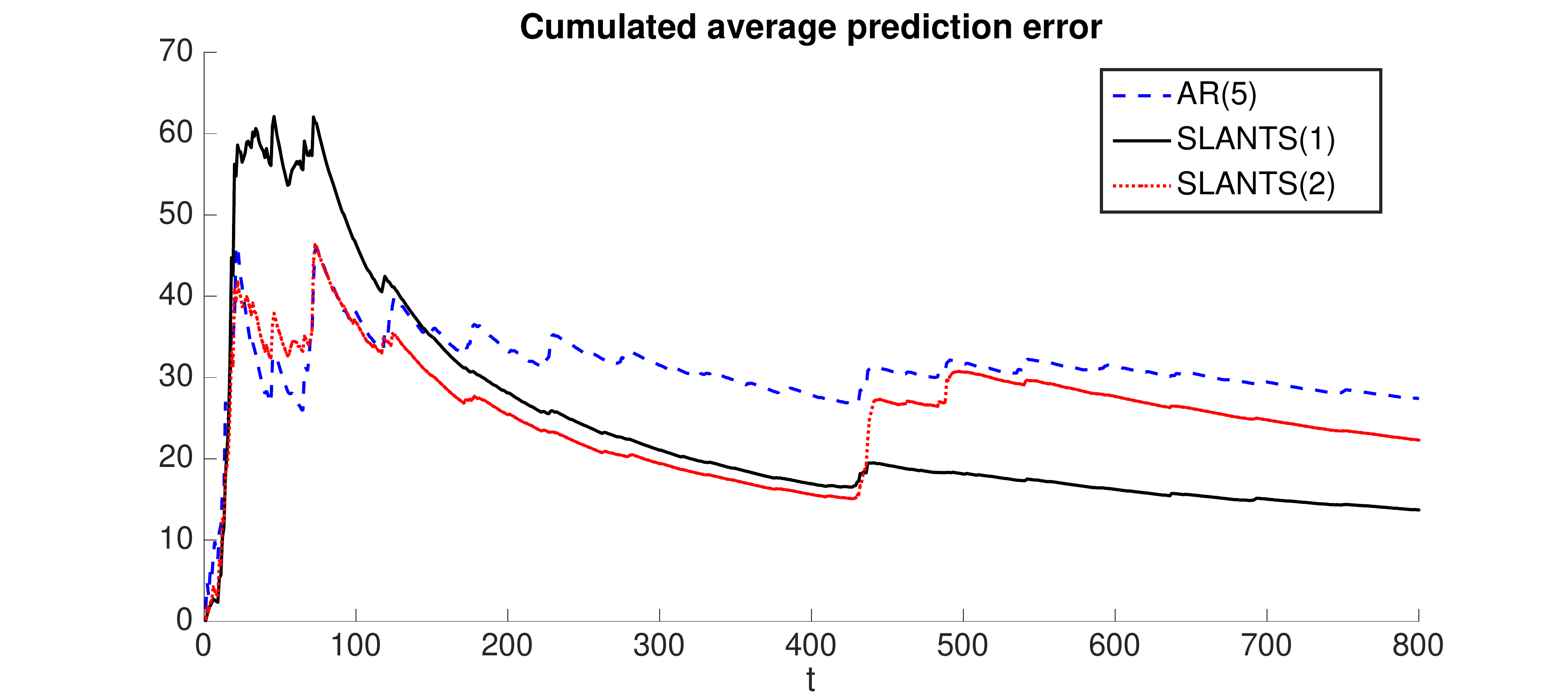}
    \end{center}
    \vspace*{-0.2in}
    \caption{A graph showing the cumulative average one-step ahead prediction error at each time step produced by three approaches: linear autoregressive model, SLANTS with linear splines, and SLANTS with quadratic splines.}
    \label{fig:unemployment}
    \vspace*{-0.2in}
    \end{figure}

Fig.~\ref{fig:unemployment} shows the cumulative average one-step ahead prediction error at each time step by the three approaches. 
Here we plot the fits to the last 800 data points due to the unstable estimates of AR and SLANTS at the beginning.
The results show that SLANTS is more flexible and reliable than linear autoregressive model in practical applications. 
Both SLANTS(1) and SLANTS(2) selected lag 1,2,52,54 as significant predictors.
It is interesting to observe that SLANTS(2) is preferred to SLANTS(1) before time step 436 (around the time when the 2008 financial crisis happened) while the simpler model SLANTS(1) is preferred after that time step. 
The fitted quadratic splines from SLANTS(2) are almost linear, which means the data has little nonlinearity.
So SLANTS(1) performs best overall.
It also demonstrates sequential LASSO as a special case of SLANTS.

\appendices
\section{Proof of Theorems}
%We start by defining some additional notations. 
We prove Theorems~1-3 in the appendix.
%\subsection{Notation}
For any real-valued column vector $x=[x_1,\ldots,x_m]$, we let $\norm{x}_2 = (\sum_{i=1}^m x_i^2)^{1/2}$, $\norm{x}_{A} = x^\T A x$ denote respectively the $\ell_2$ norm and matrix norm (with respect to $A$, a positive semidefinite matrix).
\subsection{ Proof of Theorem~\ref{thm:contraction} }
At time $T$ and iteration $k$, we define the functions $h(\cdot)$ and $g(\cdot)$ that respectively map $\hat{\bm \beta}_{T}^{(k)}$ to $\bm r_T^{(k)}$ and from $\bm r_T^{(k)}$ to $\hat{\bm \beta}_{T}^{(k+1)}$, namely $\hat{\bm \beta}_{T}^{(k)} \xrightarrow[]{h} \bm r_T^{(k)}$, $\bm r_T^{(k)} \xrightarrow[]{g} \hat{\bm \beta}_{T}^{(k+1)}$. Suppose that the largest eigenvalue of $I - \tau^2 A_{T+1}$ in absolute value is $\xi$ ($\xi < 1$). In order to prove that 
\begin{align} \label{eq:contractor}
	\norm{ g(h(\bm \chi_1)) - g(h(\bm \chi_2)) }_2 \leq \xi \norm{\bm \chi_1 - \bm \chi_2}_2 ,
\end{align}
it suffices to prove that $\norm{ h(\bm \alpha_1)-h(\bm \alpha_2) }_2 \leq \xi \norm{\bm \alpha_1-\bm \alpha_2}_2$ and  $\norm{ g(\bm \chi_1)-g(\bm \chi_2) }_2 \leq  \norm{\bm \chi_1-\bm \chi_2}_2$ for any vectors $\bm \alpha_1,\bm \alpha_2,\bm \chi_1,\bm \chi_2 $.
The first inequality follows directly from the definition of $\bm r^{(k)}$ in the E step, and $h(\bm \alpha_1)-h(\bm \alpha_2) = (I - \tau^2 A_T) (\bm \alpha_1 - \bm \alpha_2)$.  
To prove the second inequality, we prove 
\begin{align} \label{eq:cosine}
	\norm{ g(\bm \chi_{1,i})-g(\bm \chi_{2,i}) }_2 \leq  \norm{\bm \chi_{1,i} - \bm \chi_{2,i}}_2 ,
\end{align}
where $\bm \chi_{k,i}$ ($i=1,\ldots,L$) are subvectors (groups ) of corresponding to $\hat{\bm \beta}^{(k)}_{T,i}$ for either $k=1$ or $k=2$. 
For brevity we define $\tilde{\tau} = \lambda_T \tau_T^2$. 
We prove (\ref{eq:cosine}) by considering three possible cases: 1) $\norm{\bm \chi_{1,i}}_2 , \norm{\bm \chi_{2,i}}_2 \geq \tilde{\tau} $; 2) one of $\norm{\bm \chi_{1,i}}_2$ and $\norm{\bm \chi_{2,i}}_2$ is less than $\tilde{\tau}$ while the other is no less than $\tilde{\tau}$; 3)   $\norm{\bm \chi_{1,i}}_2 , \norm{\bm \chi_{2,i}}_2 < \tilde{\tau} $. 
For case 1), $g(\bm \chi_{1,i}) = g(\bm \chi_{2,i}) = \bm 0 $ and  (\ref{eq:cosine}) trivially holds. 
For case 2), assume without loss of generality that $\norm{\bm \chi_{2,i}}_2 < \tilde{\tau}$. Then 
\begin{align*}
\norm{g(\bm \chi_{1,i}) - g(\bm \chi_{2,i})}_2 &= \norm{g(\bm \chi_{1,i})}_2 
= \norm{\bm \chi_{1,i}}_2 - \tilde{\tau} 
\leq \norm{\bm \chi_{1,i}}_2 - \norm{\bm \chi_{2,i}}_2 
\leq \norm{\bm \chi_{1,i} - \bm \chi_{2,i}}_2.
\end{align*} 
For case 3), we note that $g(\bm \chi_{k,i})$ is in the same direction of $\bm \chi_{k,i}$ for $k=1,2$. We define the angle between $\bm \chi_{1,i}$ and $\bm \chi_{2,i}$ to be $\theta$, and let $a= \norm{\bm \chi_{1,i}}$, $b= \norm{\bm \chi_{2,i}}$. By the Law of Cosines, to prove $\norm{ g(\bm \chi_1)-g(\bm \chi_2) }_2^2 \leq  \norm{\bm \chi_1-\bm \chi_2}_2^2$ it suffices to prove that 
\begin{align} \label{eq:cos}
&(a-\tilde{\tau}	)^2 + (b-\tilde{\tau})^2 - 2 (a-\tilde{\tau}) (b-\tilde{\tau}) \cos (\theta) \leq a^2 + b^2 - 2 a b \cos (\theta).
\end{align}
By elementary calculations, Inequality (\ref{eq:cos}) is equivalent to 
$2 \{1-\cos(\theta)\} \{ (a+b)\tilde{\tau} - \tilde{\tau}^2)\} \geq 0$, which is straightforward. 

Finally, Inequality (\ref{eq:contractor}) and Banach Fixed Point Theorem imply that
\begin{align*}
\norm{ \hat{\bm \beta}_{T}^{(k)} - \hat{\bm \beta}_{T}}_2 \leq \frac{\xi^k}{1-\xi} \norm{ \hat{\bm \beta}_{T}^{(1)} - \hat{\bm \beta}_{T}^{(0)} }_2 
\end{align*}
which decays exponentially in $k$ for any given initial value $\hat{\bm \beta}_{T}^{(0)}$.

\subsection{Proof of Theorem~\ref{thm:step1} } %\begin{proof}
%Our proof is similar to that of \cite{huang2010variable}  for the consistency of adaptive lasso for i.i.d. covariates in high dimensional settings, and of \cite{huang2004identification}  for the consistency of BIC in finite candidate set.
The proof follows standard techniques in high-dimensional regression settings \cite{huang2010variable,hastie2015statistical}.   
  We only sketch the proof below. For brevity,  $\hat{\bm \beta}_{T}$ and $\hat{\bm \beta}_{T,d}$ are denoted as $\hat{\bm \beta}$ and $\hat{\bm \beta}_{d}$, respectively. 
  
  Let $\tilde{S}_1 = S_0 \cup S_1 $ be the set union of truly nonzero set of coefficients and the selected nonzero coefficients. By the definition of $\tilde{S}_1$, we have
  \begin{align}  
  &\norm{Y-Z_{\tilde{S}_1} \hat{\bm \beta}_{\tilde{S}_1} }_2^2 + \tilde{\lambda} \sum\limits_{d \in \tilde{S}_1} \norm{ \hat{\bm \beta}_{d} }_2   
  \leq
  \norm{Y-Z_{\tilde{S}_1} \bm \beta_{\tilde{S}_1} }_2^2 + \tilde{\lambda} \sum\limits_{d \in \tilde{S}_1} \norm{ \bm \beta_{d} }_2. \label{eq20}
  \end{align}
  Define $\bm \rho = Y - Z \bm \beta$, and $\bm \psi=Z_{\tilde{S}_1} (\hat{\bm \beta}_{\tilde{S}_1} - \bm \beta_{\tilde{S}_1}) $. 
  We obtain
  \begin{align*}
    \norm{\bm \psi}_2^2 
    &\leq 2 \bm \psi^\T \bm \rho + \tilde{\lambda} \sum\limits_{d \in \tilde{S}_1} (\norm{ \bm \beta_{d} }_2 -  \norm{ \hat{\bm \beta}_{d} }_2 ) \\
    &\leq 2 \bm \psi^\T \bm \rho + \tilde{\lambda} \sum\limits_{d \in S_0} (\norm{ \bm \beta_{d} }_2 -  \norm{ \hat{\bm \beta}_{d} }_2 ) \\
    &\leq 2 \bm \psi^\T \bm \rho + \tilde{\lambda}\sqrt{|S_0|} \norm{\bm \beta_{\tilde{S}_1} - \hat{ \bm \beta}_{\tilde{S}_1} }_2\\ 
    &\leq 2 \bm \psi^\T \bm \rho + \tilde{\lambda}\sqrt{|S_1|} \norm{\bm \beta_{\tilde{S}_1} - \hat{ \bm\beta}_{\tilde{S}_1}}_2 \\
    &\leq 2 \norm{\bm \psi}_2 \norm{\bm \rho}_2 + \tilde{\lambda}\sqrt{|S_1|} \norm{\bm \beta_{\tilde{S}_1} - \hat{ \bm\beta}_{\tilde{S}_1} }_2
  \end{align*}
  where the first inequality is rewritten from (\ref{eq20}), the second and fourth follow from $S_0 \subseteq \tilde{S}_1$, the third and fifth follow from Cauchy inequality.
  From the above equality and $2 \norm{\bm \psi}_2 \norm{\bm \rho}_2 \leq \norm{\bm \psi}_2^2/2 + 2\norm{\bm \rho}_2^2$, we obtain 
  \begin{align} 
  \norm{\bm \psi}_2^2 \leq 4 \norm{\bm \rho}_2^2 + 2 \tilde{\lambda} \sqrt{|S_1|} \norm{\bm \beta_{\tilde{S}_1} - \hat{ \bm\beta}_{\tilde{S}_1} }_2
  . \end{align}
  On the other hand, Assumption~\ref{ass:A4} gives $\norm{\bm \psi}_2^2  \geq  \kappa T \norm{\bm \beta_{\tilde{S}_1} - \hat{ \bm\beta}_{\tilde{S}_1} }_2^2 $. Therefore, 
  \begin{align*}
   \kappa T \norm{\bm \beta_{\tilde{S}_1} - \hat{ \bm\beta}_{\tilde{S}_1} }_2^2
  &\leq 4 \norm{\bm \rho}_2^2 + 2 \tilde{\lambda} \sqrt{|S_1|} \norm{\bm \beta_{\tilde{S}_1} - \hat{ \bm\beta}_{\tilde{S}_1} }_2 
  \leq 4 \norm{\bm \rho}_2^2 + \frac{2 \tilde{\lambda}^2 |S_1|}{\kappa T } + \frac{\kappa T}{2} \norm{\bm \beta_{\tilde{S}_1} - \hat{ \bm\beta}_{\tilde{S}_1} }_2^2
  \end{align*}
  which implies that 
  \begin{align} \label{eq21}
  \norm{\bm \beta_{\tilde{S}_1} - \hat{ \bm\beta}_{\tilde{S}_1} }_2^2
  \leq 8 (\kappa T)^{-1} \norm{\bm \rho}_2^2 + 4 (\kappa T)^{-2} \tilde{\lambda}^2 |S_1|  .
  \end{align}
  In order to bound $\norm{\bm \beta_{\tilde{S}_1} - \hat{ \bm\beta}_{\tilde{S}_1} }_2$, it remains to bound $\norm{\bm \rho}_2$. 
  Since $\rho_t$ can be written as 
\begin{align*}
\v_t + \sum_{d \in \tilde{S}_1} \{ f_d(X_{d,t})-f^{*}_d(X_{d,t}) \} + (\mu-\bar{Y}),
\end{align*}  
  where $(\mu-\bar{Y}) = O_p(T^{-1})$ and $\norm{ f_d - f^{*}_d }_{\infty} = O(v^{-p} + v^{1/2} T^{-1/2}) $ \cite[Lemma 1]{huang2010variable}, we  obtain 
  $\norm{\bm \rho}_2^2 \leq 2  \norm{\bm \v}_{P_X}^2 + c_2 T v^{-2p} + O_p(1)$ for sufficiently large $T$, where $c_2$ is a constant that does not depend on $v$, and
   $P_X$ is the projection matrix of $Z_{\tilde{S}_1}$. On the other side, 
\begin{align*}
\norm{\v}_{P_X}^2 \leq (\kappa T)^{-1} \norm{Z_{\tilde{S}_1}^\T \bm \v }_{2}^2 .
\end{align*}   
  Therefore, 
  \begin{align*} %\label{eq21}
  \norm{\bm \beta_{\tilde{S}_1} - \hat{ \bm\beta}_{\tilde{S}_1} }_2^2
  &\leq 8c_2 \kappa^{-1} v^{-2p}   + O(T^{-2} \norm{Z_{\tilde{S}_1}^\T \bm \v }_{2}^2 ) 
  + O_p(T^{-1}) + O(T^{-2} \tilde{\lambda}^2)
 . \end{align*}
 
 To finish the proof of (\ref{bound}), it remains to prove that $\norm{Z_{\tilde{S}_1}^\T \bm \v }_{2}^2 = O_p(T \log \tilde{D})$. Note that the elements of $\v$ are not i.i.d. conditioning on $Z_{\tilde{S}_1}$ due to time series dependency, which is different from the usual regression setting. However, for any of the $|S_1|v$ column of $Z_{\tilde{S}_1}$, say $\bm z_{d,j}$, the inner product $\bm z_{d,j}^\T \bm \v = \sum_{t=1}^T z_{d,j, t} \v_t$ is the sum  of a martingale difference sequence (MDS) with sub-exponential condition. Applying the Bernstein-type bound for a MDS, we obtain for all $w  > 0$ that
 \begin{align*}
 \P\biggl(\biggl|\sum_{t=1}^T z_{d,j, t} \v_t \biggr| > w\biggr) 
 	&\leq 2 \exp\biggl\{- \biggl(2 \sum_{t=1}^T \eta_t \biggr)^{-1} w^2  \biggr\},  \quad \textrm{ where }
 \eta_t \de var(z_{d,j, t} \v_t) \leq z_{d,j, t}^2 \sigma^2 \leq \sup_{x \in [a,b]} \{b_{d,j}(x) \}^2 \sigma^2.
 \end{align*}
 Thus, $\sum_{t=1}^T z_{d,j, t} \v_t$ is a sub-Gaussian random variable for each $d,j$.  
%
% $\P(|\sum_{t=1}^T z_{d,j, t} \v_t | > w) $ is no larger than $ 2 \exp\{-w^2 / (2 \sum_{t=1}^T \eta_t )\}$ if $0 \leq t \leq \sum_{t=1}^T \eta_t$, and no larger than $2  \exp\{-t / 2\} $ if $t > \sum_{t=1}^T \eta_t$, where $\eta_t = \sup_{z_{d,j, t}} z_{d,j, t}^2 \sigma^2 \leq \sup_{x \in [a,b]} \{b_{d,j}(x) \}^2 \sigma^2$. 
By applying similar techniques used in the maximal inequality for sub-Gaussian random variables \cite{van1996weak},  
\begin{align*}
\max_{d \in \tilde{S}_1, 1 \leq j \leq v} E( T^{-1/2} \bm z_{d,j}^\T \bm \v) \leq O(T^{-\frac{1}{2}} (\log \tilde{D})^{\frac{1}{2}}). 
\end{align*} 
Therefore,
\begin{align*}
\norm{Z_{\tilde{S}_1}^\T \bm \v }_{2}^2 &\leq |S_1|v T \max_{d \in \tilde{S}_1, 1\leq j \leq v} \{ E( T^{-1/2} \bm z_{d,j}^\T \bm \v)\}^2 
\leq  O_p(T \log \tilde{D}) .
\end{align*}
 
 To prove $\lim_{T \rightarrow \infty} \P (S_0 \subseteq S_1) = 1$, we define the event $E_{0}$ as ``There exists $d \in S_0$ such that $\hat{\bm \beta_d} = 0$ and $\bm \beta_d \neq 0$''. Under event $E_{0}$, let $d$ satisfy the above requirement. Since  
 $\norm{ f_d - f^{*}_d }_{\infty} = O(v^{-p} + v^{1/2} T^{-1/2}) $, there exists a constant $c_1^{'}$ such that for all $v \geq c_1^{'} c_0^{-1/p}$ and sufficiently large $T$, $\norm{f^{*}_d}_2 \geq c_0/2$. By a result from \cite{de1978practical}, $\norm{\bm \beta_d}_2^2 / v \geq  c_2^{'}\norm{f^{*}_d}_2^2$ holds for some constant $c_2^{'}$. Then, under $E_{0}$ it follows that  
 $\norm{\bm \beta - \hat{\bm \beta}}_2^2 \geq \norm{\bm \beta_d}_2^2 \geq c_2^{'} v c_0^2/4 \geq 16 c_2 v^{-2p}/\kappa$ for all $v \geq c_1^{''} c_0^{-2/(2p+1)}$, where $c_1^{''}$ is some positive constant. This contradicts the bound given in (\ref{bound}) for large $T$. 
%\end{proof}

\subsection{Proof of Theorem~\ref{thm:step2} } %\begin{proof}

Recall that the backward selection procedure produces a nested sequence of subsets $S_2 = \mathcal{S}^{(K)} \subseteq \cdots \subseteq \mathcal{S}^{(1)} \subseteq \mathcal{S}^{(0)} = S_1 $ 
from sequence up to $T_1$ 
 with corresponding MSE $\hat{e}^{(k)}$ ($k=0,\ldots,K$),
  where $0 \leq K \leq |S_1|-|S_2|$. 
  In addition, $\mathcal{S}^{(k)} = \mathcal{S}^{(k-1)} - \bar{d}_k^{*}$ for some $\bar{d}_k^{*} \in \mathcal{S}^{(k-1)}$ 
  using theorem 2 on the sequence.
  It suffices to prove that as $T$ goes to infinity, with probability going to one i) $S_0 \subseteq \mathcal{S}^{(k)}$ for each $k=0,\ldots,K$, and ii) $|S_2| = |S_0|$. 

Following a similar proof by \cite[Proof of Theorem 1]{huang2004identification}, it can be proved that for any $k$, 
conditioned on $S_0 \subseteq \mathcal{S}^{(k-1)} $, we have 
$\hat{e}^{(k-1)} - \hat{e}^{(k)} =  O_p(v_T /T_1) $ if   $S_0 \subseteq \mathcal{S}^{(k-1)}$, and 
$\hat{e}^{(k-1)} - \hat{e}^{(k)} = c+o_p(1) $ for some constant $c>0$ if  $S_0 \not\subseteq \mathcal{S}^{(k-1)}$.
Note that the penalty increment $ (v_T \log T_1 )/ T_1$ is larger than $O_p(v_T /T_1)$ and smaller than $c+o_p(1)$ for large $T$. %with probability going to one as $T$ goes to infinity.
By successive application of this fact finitely many times, we can prove that $S_0 \subseteq \mathcal{S}^{(k)}$ for each $k=0,\ldots,K$, and that $|S_2| = |S_0|$ with probability close to one.

%\section{}

%\section*{Acknowledgment}


\ifCLASSOPTIONcaptionsoff
  \newpage
\fi

\bibliographystyle{IEEEtran}
\balance
\bibliography{nonlinear,cp_consistency,AR_order,additional}

\end{document}